\documentclass[sigconf]{acmart} 
\usepackage[normalem]{ulem}

\usepackage{amsmath}
	\usepackage{cancel}
\usepackage{mathtools}
\usepackage{graphicx} 
\usepackage{url}
\usepackage{enumitem} 

\usepackage{stmaryrd} 

\usepackage{upgreek} 
\usepackage{bbding} 

\usepackage{listings} 
\usepackage{tabularx} 
\usepackage{booktabs} 
\usepackage{multirow}
\usepackage{hhline} 
\usepackage{diagbox}
\usepackage{pgfplotstable} 

\usepackage{xcolor,colortbl}    

\usepackage{tikz} 
\usetikzlibrary{matrix}
\usetikzlibrary{calc}
\usetikzlibrary{math}
\usetikzlibrary{arrows.meta,positioning}
\usetikzlibrary{intersections, pgfplots.fillbetween}

\usepackage{easybmat} 

\usepackage{adjustbox}
\def\eox{\unskip\kern 10pt{\unitlength1pt\linethickness{.4pt}$\diamondsuit${}}} 

\usepackage{rotating}		

\usepackage{xspace} 

\usepackage{subcaption} 
\newcommand{\hide}[1]{}

\usepackage[ruled,noend,linesnumbered]{algorithm2e} 

\usepackage{setspace} 

\DontPrintSemicolon     
\SetNlSty{}{}{}                

\SetAlFnt{\small}			
\SetAlCapFnt{\small}		
\SetAlCapNameFnt{\small}

\usepackage[capitalise]{cleveref} 

\crefname{section}{Section}{Sections}
\crefname{example}{Example}{Examples}
\crefname{figure}{Figure}{Figures}
\crefname{equation}{Equation}{Equations}
\hypersetup{                                                            
	pdfpagemode=UseNone,               
    pdfstartview=Fit,
    pdfpagelayout=SinglePage,       
    bookmarks=false,
    bookmarksnumbered=true,
    bookmarksopen=false,             
    pdftex=true,
    unicode,
    colorlinks=true,       
    linkcolor=blue,          
    citecolor=cyan,  
    filecolor=blue,      
    urlcolor=blue           
}

\usepackage{aliascnt}  	

\newaliascnt{corollary}{theorem}

\aliascntresetthe{corollary}

\newaliascnt{example}{theorem}
\newtheorem{example}[example]{Example}
\aliascntresetthe{example}

\newaliascnt{definition}{theorem}
\newtheorem{definition}[definition]{Definition}
\aliascntresetthe{definition}

\newaliascnt{proposition}{theorem}

\aliascntresetthe{proposition}

\newaliascnt{lemma}{theorem}

\aliascntresetthe{lemma}

\newaliascnt{conjecture}{theorem}

\aliascntresetthe{conjecture}

\newtheorem{questionW}{Question}
\newtheorem{resultW}{Result}

{\end{itshape}
}
\AtEndPreamble{%
    \hypersetup{colorlinks,
      linkcolor=purple,
      citecolor=blue,
      urlcolor=ACMDarkBlue,
      filecolor=ACMDarkBlue}}

\usepackage{tcolorbox}		
\tcbuselibrary{breakable,skins}		

\tcbset{examplestyle/.style={
		enhanced jigsaw,	
		colback=gray!15,	
		colframe=blue!08,	
		arc=2mm,
		boxrule=0pt,		
		left=1mm,
		right=1mm,
		left skip= 0mm,  
		right skip= 0mm, 
		top=1mm,		
		bottom=1mm,		
		breakable,		
		parbox = false,		
		before={\par\pagebreak[0]\vspace{1mm}\parindent=0pt},		
		after={\par\pagebreak[0]\vspace{1mm}\parindent=0pt},				
		bottomrule = 0mm,
		boxsep = 0mm,					
		topsep at break=0pt,			
		bottomsep at break=0pt,			
		pad at break=0mm,
		pad before break=1mm,		
		pad after break=1mm,		
		bottomrule at break=0mm,
		toprule at break=0mm,		
		}}

\usepackage{footnote}

\tcolorboxenvironment{example}{examplestyle}

\makeatletter
\DeclareRobustCommand*\uell{\mathpalette\@uell\relax}
\newcommand*\@uell[2]{
  \setbox0=\hbox{$#1\ell$}
  \setbox1=\hbox{\rotatebox{10}{$#1\ell$}}
  \dimen0=\wd0 \advance\dimen0 by -\wd1 \divide\dimen0 by 2
  \mathord{\lower 0.1ex \hbox{\kern\dimen0\unhbox1\kern\dimen0}}
}
\makeatother

\usepackage{marginnote}

\renewcommand{\epsilon}{\varepsilon} 

\usepackage{scalerel}
\usepackage{tikz}
\usetikzlibrary{svg.path}

\definecolor{orcidlogocol}{HTML}{A6CE39}
\tikzset{
  orcidlogo/.pic={
    \fill[orcidlogocol] svg{M256,128c0,70.7-57.3,128-128,128C57.3,256,0,198.7,0,128C0,57.3,57.3,0,128,0C198.7,0,256,57.3,256,128z};
    \fill[white] svg{M86.3,186.2H70.9V79.1h15.4v48.4V186.2z}
                 svg{M108.9,79.1h41.6c39.6,0,57,28.3,57,53.6c0,27.5-21.5,53.6-56.8,53.6h-41.8V79.1z M124.3,172.4h24.5c34.9,0,42.9-26.5,42.9-39.7c0-21.5-13.7-39.7-43.7-39.7h-23.7V172.4z}
                 svg{M88.7,56.8c0,5.5-4.5,10.1-10.1,10.1c-5.6,0-10.1-4.6-10.1-10.1c0-5.6,4.5-10.1,10.1-10.1C84.2,46.7,88.7,51.3,88.7,56.8z};
  }
}

\RequirePackage{etoolbox}
\DeclareRobustCommand\orcidicon[1]{\href{https://orcid.org/#1}{\mbox{\scalerel*{
\begin{tikzpicture}[yscale=-1,transform shape]
\pic{orcidlogo};
\end{tikzpicture}
}{|}}}}

\usepackage{xcolor}
\usepackage{ifthen}
\usepackage{xfp}

\definecolor{mygreen}{rgb}{0.439, 0.678, 0.278}
\definecolor{myred}{rgb}{1, 0, 0}

\newcommand{\colordelta}[1]{%
  \ifthenelse{\lengthtest{#1pt > 0pt}}%
    {( \textcolor{mygreen}{#1} )}%
    {( \textcolor{myred}{#1} )}%
}
\newcommand{\colordeltanp}[1]{%
  \ifthenelse{\lengthtest{#1pt > 0pt}}%
    {\textcolor{mygreen}{#1}}%
    {\textcolor{myred}{#1}}%
}

\newcommand{\eat}[1]{}

\def\header{\vspace{0.5mm} \noindent}

\newcommand{\model}{{REVEAL}\xspace}
\newcommand{\modelplus}{{REVEAL+}\xspace}

\newcommand{\turl}{{TURL}\xspace}
\newcommand{\sato}{{Sato}\xspace}
\newcommand{\doduo}{{Doduo}\xspace}
\newcommand{\starmie}{Starmie\xspace}
\newcommand{\watchog}{Watchog\xspace}
\newcommand{\tllama}{TableLlama\xspace}
\newcommand{\qwen}{Qwen-Plus\xspace}

\newcommand{\reca}{{RECA}\xspace}

\newcommand{\qscore}{quality score\xspace}
\newcommand{\colcontext}{column context\xspace}
\newcommand{\colcontexts}{column contexts\xspace}

\newcommand{\datasetvt}{\mathcal{D}_{\mathrm{v}}^{\mathrm{train}}}
\newcommand{\datasett}{\mathcal{D}^{\mathrm{train}}}
\newcommand{\datasetp}{\mathcal{D}^{\mathrm{train}}_{\mathrm{pair}}}

\newcommand{\micro}{Micro-F1\xspace}
\newcommand{\macro}{Macro-F1\xspace}
\newcommand{\gitdb}{GitTablesDB\xspace}
\newcommand{\gitsc}{GitTablesSC\xspace}
\newcommand{\sotabcta}{SOTAB-CTA\xspace}
\newcommand{\sotabcpa}{SOTAB-CPA\xspace}
\newcommand{\wikicta}{WikiTables-CTA\xspace}
\newcommand{\wikicpa}{WikiTables-CPA\xspace}

\newcommand{\target}{\tau\xspace}
\newcommand{\candpool}{\mathcal{C}\xspace}
\newcommand{\vpool}{\mathcal{S}\xspace}
\newcommand{\vmodel}{\Pi\xspace}
\newcommand{\tmodel}{f\xspace} 
\newcommand{\embcol}{\mathbf{e}_c\xspace}
\newcommand{\mr}{g\xspace}
\newcommand{\gt}{gt_{\target}\xspace}
\newcommand{\vcan}{\mathcal{S}'\xspace} 
\newcommand{\vlabel}{{y_{\vcan}^{\target}\xspace}}

\makeatletter
\def\@ACM@checkaffil{
    \if@ACM@instpresent\else
    \ClassWarningNoLine{\@classname}{No institution present for an affiliation}%
    \fi
    \if@ACM@citypresent\else
    \ClassWarningNoLine{\@classname}{No city present for an affiliation}%
    \fi
    \if@ACM@countrypresent\else
        \ClassWarningNoLine{\@classname}{No country present for an affiliation}%
    \fi
}
\makeatother
\AtBeginDocument{%
  }

\setcopyright{acmlicensed}
\copyrightyear{2025}
\acmYear{2026}
\acmDOI{XXXXXXX.XXXXXXX}
\acmConference[SIGMOD '26]{Proceedings of the 2026 ACM SIGMOD International Conference on Management of Data (SIGMOD '26)}{May 31--June 05, 2026}{Bengaluru, India}

\acmISBN{978-1-4503-XXXX-X/2018/06}

\begin{document}

\title{Retrieve-and-Verify: A Table Context Selection Framework for Accurate Column Annotations}

\author{Zhihao Ding}
\authornote{Co-primary authors.}
\affiliation{%
  \institution{Hong Kong Polytechnic University}
}
\email{tommy-zh.ding@connect.polyu.hk}

\author{Yongkang Sun}
\authornotemark[1]
\affiliation{%
  \institution{Hong Kong Polytechnic University}
}
\email{yongkang.sun@connect.polyu.hk}

\author{Jieming Shi}
\affiliation{%
  \institution{Hong Kong Polytechnic University}
}
\email{jieming.shi@polyu.edu.hk}
\renewcommand{\shortauthors}{Ding, Sun, and Shi}

\begin{abstract}
Tables are a prevalent format for structured data, yet their metadata, such as semantic types and column relationships, is often incomplete or ambiguous. Column annotation tasks, including Column Type Annotation (CTA) and Column Property Annotation (CPA),  address this by leveraging table context, which are critical for data management. Existing methods typically serialize all columns in a table into pretrained language models to incorporate context, but this coarse-grained approach often degrades performance in wide tables with many irrelevant or misleading columns. 
To address this, we propose a novel retrieve-and-verify context selection framework for accurate column annotation, introducing two methods: \model and \modelplus. 
In \model, we design an efficient unsupervised retrieval technique to select compact, informative column contexts by balancing semantic relevance and diversity, and develop context-aware encoding techniques with role embeddings and target-context pair training to effectively differentiate target and context columns. To further improve performance, in \modelplus, we design a verification model that refines the selected context by directly estimating its quality for specific annotation tasks. To achieve this, we formulate a novel column context verification problem as a classification task and then develop the verification model.  Moreover, in \modelplus, we develop a top-down verification inference technique to ensure efficiency by reducing the search space for high-quality context subsets from exponential to quadratic.
Extensive experiments on six benchmark datasets demonstrate that our methods consistently outperform state-of-the-art baselines. 
\end{abstract}

\maketitle

\section{Introduction}
\label{sec:intro}

Relational tables are a fundamental format for organizing structured data in diverse applications. As structured data grows in volume and complexity, understanding tables through metadata becomes critical for efficient management and utilization~\cite{DBLP:conf/www/BrickleyBN19,doduo}. 
However, table metadata, particularly the semantic types and relationships of columns, is often missing, incomplete, or ambiguous~\cite{DBLP:conf/esws/Jimenez-RuizHEC20,DBLP:journals/tods/LeventidisRGMR23}. 
Column annotation mitigates this via important tasks, including
Column Type Annotation (CTA), which predicts a target column's semantic type, and Column Property Annotation (CPA), also known as Column Relationship Annotation, which identifies relationships of column pairs. Accurate column annotation is crucial for applications such as schema matching~\cite{DBLP:conf/icde/KoutrasSIPBFLBK21, DBLP:conf/icde/DuYW0L24}, dataset discovery~\cite{starmie, santos}, data integration~\cite{DBLP:journals/pacmmod/TuFTWL0JG23,DBLP:journals/pvldb/KhatiwadaSGM22}, and semantic data versioning~\cite{DBLP:journals/pvldb/ShragaM23}. 

\begin{figure}[t]
  \centering
    \includegraphics[width=\linewidth]{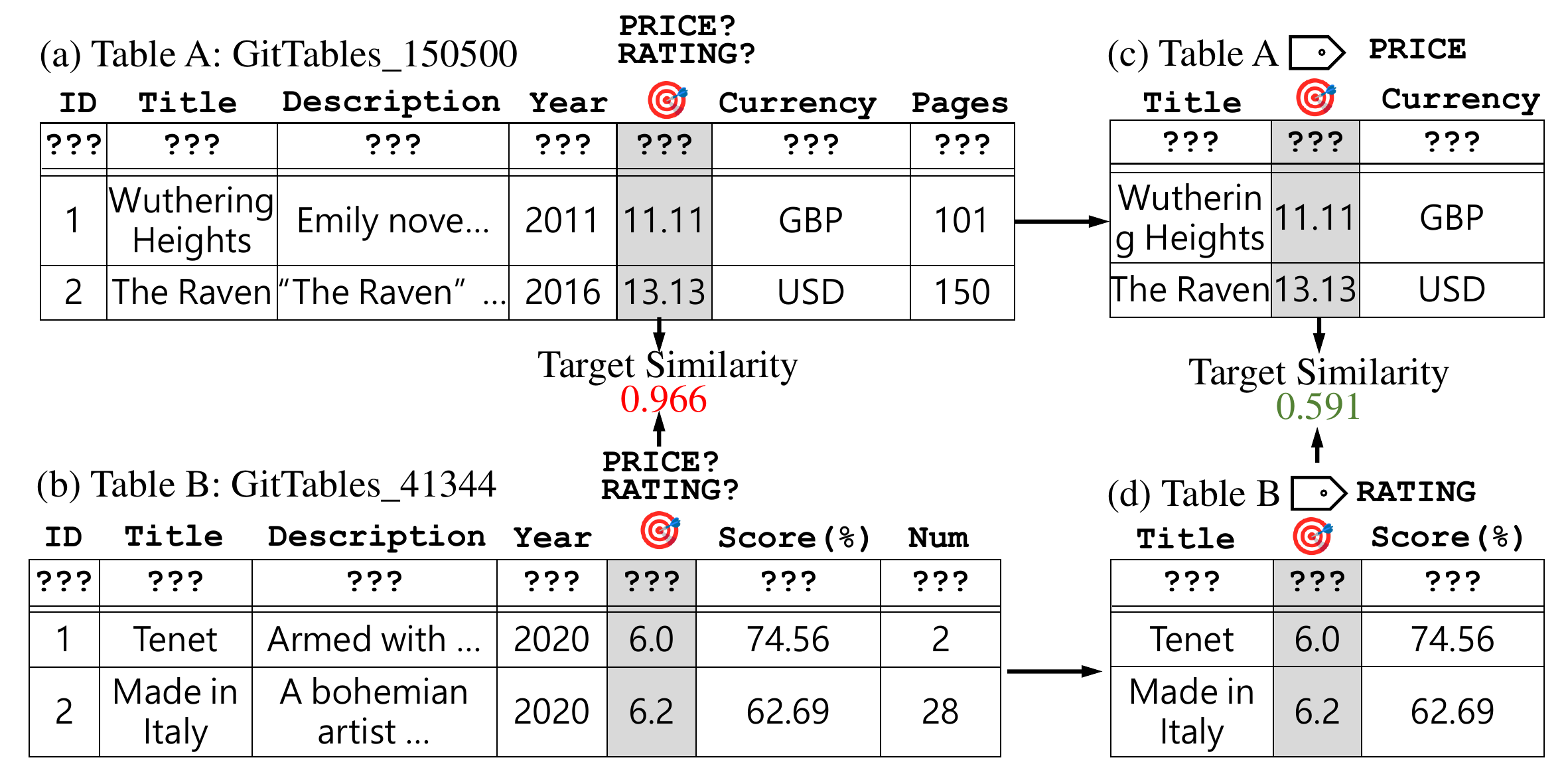}
  \caption{Example on two real-world tables: (a)(c) Raw tables; (b)(d) Tables with selected contexts.}
  \label{fig:example2}
  \vspace{-3mm}
\end{figure}

\begin{figure}[t]  
  \centering
\includegraphics[width=0.82\linewidth]{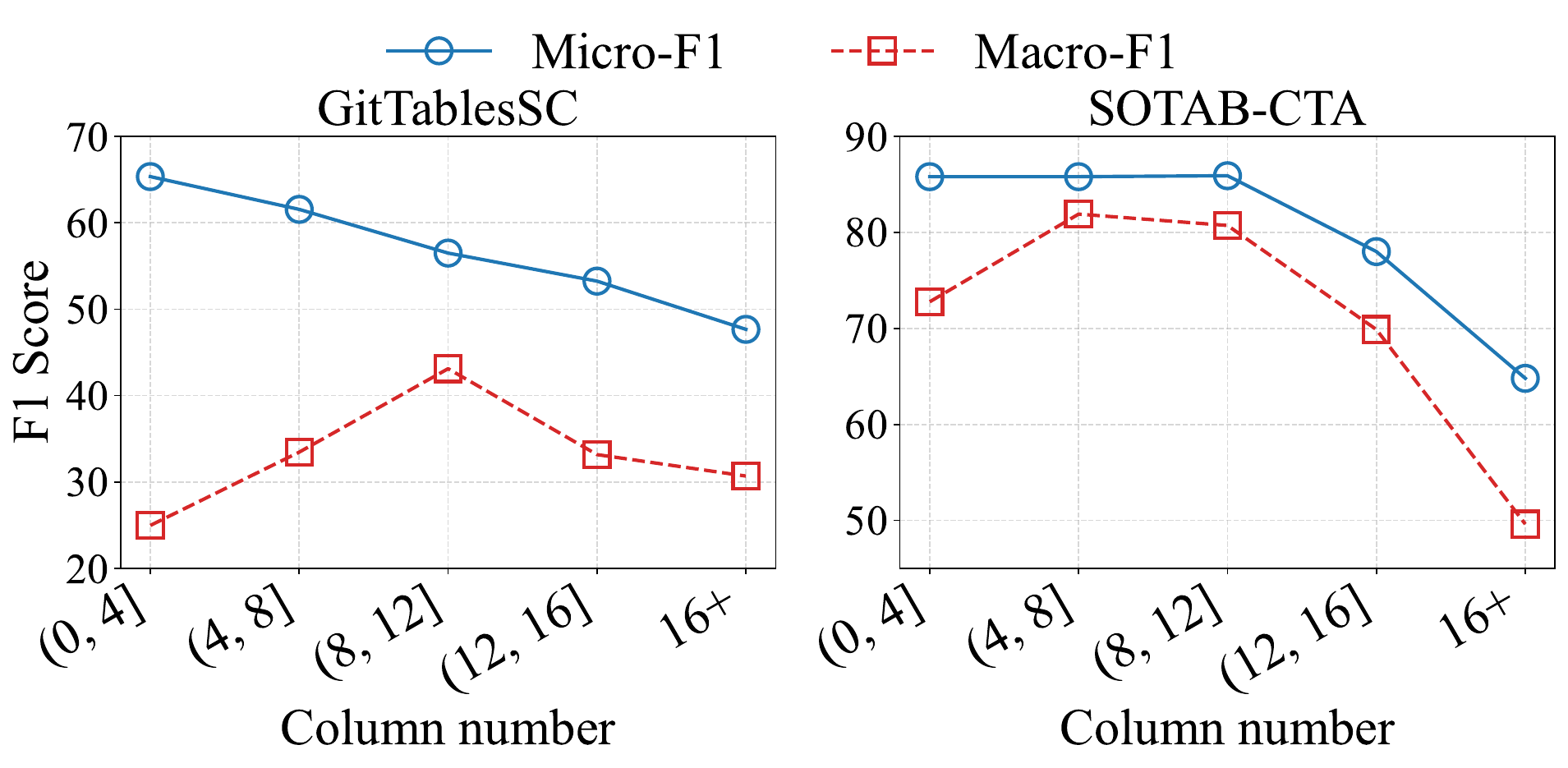}
      \caption{The impact of column number on CTA.}
  \label{fig:doduo}
  \vspace{2mm}
\end{figure}

In this work, we investigate CTA and CPA in scenarios where metadata is unavailable. Accurate annotation of a target's semantic type or relationship requires the context of other columns within the same table.
For example, in Table A of \cref{fig:example2}(a), the target column (in gray) contains float values such as `11.11' and `13.13'. 
Solely relying on the target itself, it is difficult to determine if it represents \texttt{PRICE} or \texttt{RATING} semantic type.
However, when  considering the other columns, the target is likely to be \texttt{PRICE} since it is followed by a currency column, and the table describes books.

Recent studies~\cite{turl,doduo,reca,watchog} have leveraged language models (LMs) such as BERT~\cite{bert} and LLaMA~\cite{llama3} to enhance column annotation by utilizing column representations as context~\cite{sato,doduo,starmie}, and even inter-table columns~\cite{reca}. 
A common approach is to serialize all columns in a table into a single sequence and input it into LMs, such as BERT in \doduo~\cite{doduo}, to generate column embeddings that capture patterns and relationships between columns.  
Then the model predicts the semantic type or relationship of the target as a classification problem based on the generated embeddings.

While existing methods have made notable progress, their approach to leveraging column context remains coarse-grained. This often leads to sub-optimal performance, particularly in wide tables with many columns, which are common in real-world scenarios~\cite{tablegpt}.
We conduct an empirical analysis of \doduo on two datasets, \gitsc and \sotabcta, to examine the impact of the number of columns in a table on model performance (see detailed experiments in \cref{sec:exp}). As shown in \cref{fig:doduo}, performance improves in terms of \macro and \micro as the number of columns increases from 1 to 12, demonstrating the benefit of leveraging additional column content. However, performance declines when the number of columns exceeds 12. This indicates that wide tables with many columns pose greater challenges, as not all columns are beneficial for the target. Including all columns may introduce noise and redundancy,  degrading performance~\cite{DBLP:conf/iclr/YoranWRB24,DBLP:conf/icml/ShiCMSDCSZ23}.
We provide \cref{example:intro} with two real tables in  \cref{fig:example2} to illustrate this issue.

\vspace{1mm}
\begin{example} \label{example:intro} 
  \vspace{1mm}
\cref{fig:example2}(a) and (c) show two real tables with target columns in grey and the goal is to predict their semantic types, either \texttt{PRICE} or \texttt{RATING}, which are shown for illustration but not available during annotation. Both targets contain float values, and the tables include columns with similar data types, such as text and numbers. Using all columns as context, a model may fail to distinguish the targets, leading to incorrect predictions. For instance, \doduo generates embeddings for the two targets with a high cosine similarity of 0.966, making them hard to differentiate. 

However, closer inspection reveals that the target in Table A is likely \texttt{PRICE}, as it is followed by a currency column with GBP and USD values, while the target in Table B is likely \texttt{RATING}, as it is followed by a percentage column. Selecting only relevant columns, such as \texttt{TITLE} and \texttt{CURRENCY} for Table A, and \texttt{TITLE} and \texttt{SCORE(\%)} for Table B, as shown in \cref{fig:example2}(b) and (d), reduces the cosine similarity between the targets to 0.591, making them more distinguishable. This highlights the importance of selecting relevant context columns for accurate annotation.
\vspace{1mm}
\end{example}

These empirical findings reveal the importance of explicitly selecting relevant column contexts for accurate column annotation.
To this end, we propose a novel retrieve-and-verify context selection framework comprising the \model and \modelplus methods. 
\model delivers superior annotation performance with high efficiency, while \modelplus further enhances annotation accuracy significantly with moderate additional overhead.

In \model, we first design an efficient \textit{retrieval method} to select a compact, informative subset of columns from the input table $T$ as the \textit{column context} $\candpool$ for a given target in an unsupervised manner (\cref{sec:retrieval}). This method ensures that $\candpool$ is both semantically relevant and diverse, providing meaningful context without requiring labeled supervision. 
Next, to generate high-quality embeddings that effectively distinguish target and context columns, we develop \textit{context-aware encoding techniques} (\cref{sec::encoding}). This includes a context-aware encoder with role embeddings to explicitly mark column roles, and a target-context pair training strategy that treats each target-context pair as a distinct training unit, enabling effective modeling of target-context interactions.

To further improve performance, we extend \model with \modelplus, incorporating a \textit{verification model} to refine $\candpool$ into a verified column context $\vpool$. 
The verification directly estimates the quality of the selected column context for the target in a supervised manner, ensuring that the context is not only relevant but also effective for the specific annotation tasks. 
To this end, we formulate a novel \textit{column context verification} problem as a classification task, construct a pseudo-labeled dataset, and design and train the verification model (\cref{sec::verification_model}). 
To avoid the exponential search space of all subsets of $\candpool$ to find the highest quality $\vpool$, we develop a \textit{top-down verification inference method} that efficiently obtains $\vpool$ using a greedy strategy, reducing the search space to quadratic (\cref{sec::inference}).

\eat{\begin{example} \label{example:3} Consider the two tables in \cref{fig:example2}(b). For the target columns in Table A and Table B, selecting relevant contextual columns such as \texttt{TITLE} and \texttt{CURRENCY} for Table A, and \texttt{TITLE} and \texttt{SCORE(\%)} for Table B, provides sufficient information to infer column types while avoiding ambiguity from irrelevant columns. With this curated context, the embedding similarity between the two target columns decreases from 0.966 to 0.591, making their semantics more distinguishable and facilitate correct prediction. \end{example}}

We conduct extensive experiments on six benchmark datasets, comparing our methods with state-of-the-art baselines. The results show that \model and \modelplus consistently outperform existing approaches, demonstrating that our techniques in the proposed context selection framework advance column annotation performance.

We summarize our contributions below:
\vspace{-\topsep+3pt}
 \begin{itemize}[leftmargin=*]
    \item We propose a retrieve-and-verify column context selection framework, introducing the \model and \modelplus methods to address the challenges of noisy and redundant column contexts in column annotation tasks.
    \item We propose a retrieval method to select compact and informative column contexts and introduce context-aware encoding to differentiate target and context columns for better embeddings.
    \item We design a verification model with a novel problem formulation and an efficient top-down inference strategy to refine the selected column context, further boosting annotation performance.
    \item Extensive experiments on six benchmark datasets validate the effectiveness and efficiency of our proposed methods.
 \end{itemize}

\begin{table}[t]
    \centering
    \caption{Frequently used notations.}
    \label{tab:notations}
    \small
    \vspace{-3.5mm}
    \resizebox{0.94\linewidth}{!}{
          \renewcommand{\arraystretch}{0.98}
        \begin{tabular}{ll}
        \toprule
        \textbf{Notation} & \textbf{Description} \\
        \midrule
        $T$ & A table with multiple columns. \\
        $c_i$     & The $i$-th column in the table. \\
        $\target$ & The target column or column pair for annotation. \\
        $\gt$     & The ground-truth of the target. \\
        $\candpool$ & The retrieved column context for $\target$. \\
        $\vpool$ & The verified column context for $\target$. \\
        $\vcan$ & A subset of columns from $\candpool$. \\ 
        $\vmodel$ & The verification model. \\
        $\vmodel(\target, \vpool)$ & The \qscore assigned by the verification model. \\

        $\tmodel$ & The prediction module. \\
        $\mathbf{h}^{\target}_{\vcan}$ & The embedding of $\target$ with context $\vcan$. \\
        $\vlabel$ & The label indicating the quality of a column context. \\
        $K$ & The desired size of the retrieved column context $\candpool$. \\
        \bottomrule
    \end{tabular}
    }
    \vspace{1mm}
\end{table}

\begin{figure*}[t!] 
    \centering
      \includegraphics[width=0.90\textwidth]{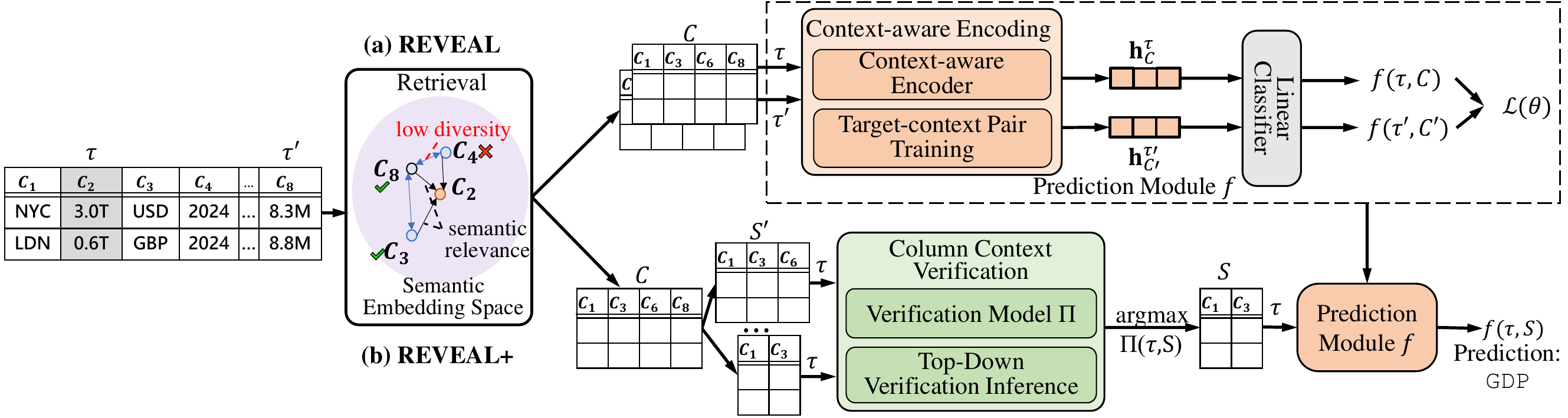}
     \vspace{1mm}
     \caption{The Framework of \model and \modelplus.}
   
    \label{fig:framework}
    \vspace{-1mm}
\end{figure*}

\section{Preliminaries}
\label{sec:preliminaries}

We focus on two column annotation tasks: {Column Type
Annotation (CTA)} and {Column Property Annotation (CPA)}~\cite{sato,doduo,turl,watchog}. As illustrated in \cref{fig:example2}, 
CTA assigns semantic types (e.g., \texttt{PRICE}, \texttt{RATING}) to individual columns, providing a clearer understanding of their semantics.
CPA  aims to determine the semantic relationship between column pairs, such as \texttt{place\_of\_birth} linking \texttt{person} to \texttt{city}, or \texttt{has\_population} linking \texttt{city} to \texttt{population}.
Instead of primitive data types (e.g., {String}, {Int}, {Timestamp}), CTA and CPA focus on annotating columns with semantic types and relationships, offering deeper semantic understanding of tabular data.
Formally, a  table $T$ with $n$ columns is represented as 
$T = (c_1, c_2, \ldots, c_n)$, where $c_i$ denotes the $i$-th column in the table. Each column $c_i$ comprises $m$ cell values, expressed as $c_i = (v_1^i, v_2^i, \ldots, v_m^i)$, where $v_j^i$ represents the value of the $j$-th cell in column $c_i$. 
The definitions of CTA and CPA~\cite{watchog,doduo} are as follows:

\begin{definition}[Column Type Annotation (CTA)]
\label{def:cta}
Given a table $T$ with $n$ columns, and a target column $\target = c_i$, where $1 \leq i \leq n$, and a set of possible semantic types $\mathcal{T}$, 
the task of CTA is to develop  a model {$\mathcal{M}$} that can predict a type label $\mathcal{M}(\target, T)\in\mathcal{T}$
so that each cell in $\target$ has the same semantic type. 
\end{definition}

\begin{definition}[Column Property Annotation (CPA)]
\label{def:cpa}
Given a table $T$ with $n$ columns, and a target column pair $\target = (c_i, c_j)$, where $1 \leq i,j \leq n$, and a set of possible relation types $\mathcal{R}$, the task of CPA is to develop a model {$\mathcal{M}$} that can predict a relation type $\mathcal{M}(\target, T)\in\mathcal{R}$ to represent the relationship between the two columns.
\end{definition}

We focus on the setting where no metadata, such as headers or captions, is available, as these are often missing in many datasets~\cite{gittables,sotab}.
Hereafter, when the context is clear, we use target $\target$ to denote either a   target column in CTA or a pair of target columns in CPA.

\cref{tab:notations} summarizes the frequently used notations in this paper.

\section{Overview}
\label{sec:method}

We provide an overview of our methods \model and \modelplus. Our methods are designed to be applicable to both CTA and CPA. We mainly explain our methods in the context of CTA for clarity.

\cref{fig:framework}(a) illustrates  the \model method. 
Given a table $T$ with $n$ columns, there may be different targets, e.g., $\target$ and $\target'$, to be annotated.
As discussed, treating all columns in $T$ as a shared context for different targets can lead to degraded performance. 
Therefore, in \model, we first design a {retrieval method} to explicitly select a subset of columns from $T$ as the \textit{column context} $\candpool$ for a target $\target$, considering semantic relevance and diversity among columns in $T$
 (\cref{sec:retrieval}). 
The retrieval method operates solely on unsupervised information, ensuring scalability and effectiveness. 
Then the next step is to encode them into embeddings via language models.
Existing studies~\cite{doduo,watchog} typically serialize all columns into a sequence, treating the target and context columns uniformly. 
However, we argue that the embedding of a column should adapt based on its role—whether as the target or as part of the context.
To achieve this, we develop {context-aware encoding techniques} that explicitly differentiate the target from its context columns in embeddings (\cref{sec::encoding}). Specifically, we introduce a context-aware encoder with role embeddings to mark the roles of columns during the embedding process, and we adopt a target-context pair training strategy, where each target $\target$ is paired with its column context $\candpool$ as distinct training units. The obtained embeddings   are then passed to a linear layer for generating the final predictions.
 
\model excels state-of-the-art methods in effectiveness and efficiency, as shown in the experiments. While efficiency is important, annotation quality often takes precedence, especially for offline tasks. To further enhance annotation quality, \modelplus improves effectiveness with moderate additional computational cost.

\modelplus is shown in \cref{fig:framework}(b). Similar to \model, it starts with the same retrieval method to obtain   $\candpool$ for the target $\target$. While $\candpool$ is selected based on unsupervised information in $T$, the columns in $\candpool$ are not yet \textit{verified} whether they are truly helpful in the annotation tasks.
To address this, \modelplus incorporates a {verification model} (\cref{sec:verification}) to further refine $\candpool$ to obtain a verified column context $\vpool$ for the target $\target$, which retains only the most informative columns for the target $\target$ on the annotation tasks.
To improve the efficiency of obtaining $\vpool$, we propose a {top-down verification inference} method (\cref{sec::inference}).
As formulated in \cref{sec::verification_model}, the verification model $\vmodel$ evaluates a subset $\vcan$ of $\candpool$ for a target $\target$ by outputting a \qscore $\vmodel(\target, \vcan)$, which represents the likelihood that $\vcan$ enables the correct annotation of $\target$. This verification task is framed as a classification problem: determine whether a subset $\vcan$ of $\candpool$ is effective for accurately annotating the target $\target$.
The subset of $\candpool$ with the maximum quality score is selected as the verified column context $\vpool$ for the target $\target$.
A key challenge is the lack of labeled data for training the verification model. 
To address this, we construct a labeled dataset by leveraging the predictions of the trained prediction module $\tmodel$ on the training and validation datasets, which are available after $\tmodel$ is trained.
For each subset $\vcan \subseteq \candpool$, we label it as positive if $\tmodel$ makes a correct prediction for $\target$ using $\vcan$ as context, indicating that $\vcan$ is a good context. Otherwise, it is labeled as negative. 
The model $\vmodel$ is then trained on this constructed labeled dataset.
During inference, the model $\vmodel$ finds the   $\vpool$ that maximizes the   score $\vmodel(\target, \vcan)$ for  $\target$. The trained prediction module $\tmodel$ then uses $\vpool$ to generate the final prediction for $\target$. Evaluating all subsets $\vcan \subseteq \candpool$ is infeasible due to the exponential search space of $2^{|\candpool|}$. Hence, we propose a top-down verification inference method (\cref{sec::inference}) that iteratively refines $\candpool$, reducing the complexity to $O(|\candpool|^2)$, with early stopping to improve efficiency.

\section{The REVEAL Method}\label{sec:reveal}

\subsection{Column Context Retrieval} 
\label{sec:retrieval}

As mentioned, in a wide table $T=(c_1, \ldots, c_n)$, using all columns for annotating a target $\target$ can degrade performance due to irrelevant columns. We aim to retrieve a subset of columns as the \colcontext $\candpool$ for $\target$. However, the search space of $2^n$ subsets grows exponentially with $n$, and different targets may require distinct \colcontexts.
In this section, we propose an efficient retrieval method that leverages only the unsupervised information of columns in $T$ to retrieve a compact yet informative \colcontext $\candpool$ for $\target$.

\header\textbf{Intuitions.}
As shown in \cref{fig:example2}, not all columns in a table equally contribute to understanding the target $\target$ with ground truth  \texttt{PRICE}. For example, the \texttt{Description} column, despite containing detailed text, is irrelevant to the target's semantics, while the \texttt{CURRENCY} column provides essential context. 
 Including irrelevant columns can degrade performance, highlighting the need to filter for semantically relevant columns. 
One naive approach is to select the top-$K$ most similar columns to  $\target$ as $\candpool$. However, this may not always yield an informative context. For example, consider constructing a \colcontext $\candpool$ of size 3 for the target  in \cref{fig:example2}(a). The target $\target$ contains numerical values, and the cosine similarity between the target and other columns are: \texttt{ID} (0.82), \texttt{PAGES} (0.76), \texttt{YEAR} (0.68), \texttt{CURRENCY} (0.55), \texttt{TITLE} (0.28), and \texttt{DESCRIPTION} (0.21). Selecting the top-3 most similar columns would yield \texttt{ID}, \texttt{PAGES}, and \texttt{YEAR}. However, these columns are all numerically related and fail to provide diverse contextual information. In contrast, \texttt{CURRENCY} (0.55), though less similar, specifies the nature of the values in the target and is crucial for inferring its semantic type. Thus, a more balanced approach that considers both \textit{similarity and diversity} is needed.

\header\textbf{Retrieval Method.}
To select the \colcontext $\candpool$ of a target $\target$ in a table $T$, we consider both semantic similarity and diversity among columns. 
First, each column $c$ in $T$ is serialized into a string by concatenating its cell values $v_1, \ldots, v_m$. 
This serialized string is then passed through a text encoder $\mathcal{E}$~\cite{DBLP:conf/emnlp/ReimersG19}, which generates a dense embedding $\embcol$ for the column. 
The embeddings of semantically similar columns are closer in the embedding space.
\begin{equation}\label{eq:text}
    \small
    \embcol = \mathcal{E}(\mathrm{CONCAT}(v_1, \ldots, v_m)), 
\end{equation}
where $\mathrm{CONCAT}(\cdot)$ represents the concatenation operation, $\mathcal{E}(\cdot)$ is the text encoder, $\embcol \in \mathbb{R}^{d_e}$ denotes the embedding of column $c \in T$, and $d_e$ is the embedding dimension.

\SetKwInOut{Parameter}{Variables}
\begin{algorithm}[t]
	\caption{Column Context Retrieval}
	\label{alg:retrieval}
\small
	\KwIn{Table $T$, target $\target$,  encoder $ \mathcal{E}$, size $K$}
	\KwOut{Column context $\candpool$}

	\ForEach{$c \in T$}{
	    $\embcol \gets \mathcal{E}(c)$ \tcp{Encode columns}
	}

	$\candpool \gets \emptyset$
	
	$c' \gets \underset{c \in T \setminus \{\target\}}{\operatorname{argmax}} \operatorname{cos}(\mathbf{e}_{c}, \mathbf{e}_{\target})$ \tcp{Select the first column} 
	
	$\candpool \gets \candpool \cup \{c'\}$

    \tcp{Iterative selection} 
	\While{$|\candpool| < K$}
    {
	    $c' \gets \underset{c \in T \setminus \candpool}{\operatorname{argmax}} \left[ \operatorname{cos}(\mathbf{e}_{c}, \mathbf{e}_{\target}) - \max_{c'' \in \candpool} \operatorname{cos}(\mathbf{e}_{c}, \mathbf{e}_{c''}) \right]$
	    
	    $\candpool \gets \candpool \cup \{c'\}$
	}
	
	\Return $\candpool$
\end{algorithm}

Then we use maximal marginal relevance~\cite{DBLP:conf/sigir/CarbonellG98} as a measure to balance relevance and diversity, and develop an iterative process to construct $\candpool$. 
Starting with an empty $\candpool=\emptyset$, we iteratively select the column $c \in T \setminus \candpool$ that maximizes the marginal relevance score $\mr(c,\target, \candpool)$, and add it to $\candpool$. 
In \cref{eq:candpoolGeneration}, $\mr(c,\target, \candpool)$ is defined as the difference between the semantic similarity of $c$ to the target $\target$, $cos(\embcol, \mathbf{e}_{\target})$, and the maximum similarity of $c$ to any column already in $\candpool$, $\max_{c'' \in \candpool } cos(\embcol, \mathbf{e}_{c''})$. 
The first term ensures that the selected column is relevant to the target, while the second term penalizes redundancy by discouraging columns similar to those already in $\candpool$. 
This process continues until $\candpool$ reaches the desired size $K$. 
\begin{equation} \label{eq:candpoolGeneration}
    \small
    \begin{split}
    \mr(c,\target, \candpool) & =   \operatorname{cos}(\embcol, \mathbf{e}_{\target}) -   \max_{c'' \in \candpool } \operatorname{cos}(\embcol, \mathbf{e}_{c''}) \\
    c' &= \underset{c \in T \setminus \candpool }{\operatorname{argmax }} \text{ }\mr(c,\target, \candpool).
    \end{split}
    \end{equation}

\begin{example}
    \label{example:diversity} 
In \cref{fig:example2}(b), for a target $\target$ containing numerical values, selecting the top-3 most similar columns retrieves other numeric columns (e.g., \texttt{ID}, \texttt{YEAR}, \texttt{PAGES}), which provide limited assistance in clarifying the true semantic meaning of the target, i.e., \texttt{PRICE}.
In contrast, our retrieval method selects a more semantically meaningful size-3 \colcontext $\candpool$.
We initialize $\candpool$ as empty and iteratively add columns based on their marginal relevance scores.
First, the most similar column to the target, \texttt{ID} (0.82), is added, resulting in $\candpool = \{\texttt{ID}\}$.
Next, we compute $\mr(c,\target,\candpool)$ for each column $c \in T \setminus \candpool$. For instance, \texttt{PAGES} has a high similarity to the target (0.76) but also a high similarity to \texttt{ID} (0.71), yielding $\mr(\texttt{PAGES},\target,\candpool) = 0.05$. On the other hand, \texttt{CURRENCY} has $cos(\embcol, \mathbf{e}_{\target}) = 0.55$ and $cos(\embcol, \mathbf{e}_{\texttt{ID}}) = 0.34$, resulting in $\mr(\texttt{CURRENCY},\target,\candpool) = 0.21$, making it more favorable than \texttt{PAGES}.
Similarly, other columns in $T \setminus \candpool$ have lower marginal relevance scores than \texttt{CURRENCY}, so we expand $\candpool$ to $\{\texttt{ID}, \texttt{CURRENCY}\}$.
Finally, we repeat the process to select the third column. \texttt{TITLE} achieves the highest marginal relevance score, surpassing \texttt{PAGES}, \texttt{YEAR}, and \texttt{DESCRIPTION}. 
Thus, the final $\candpool$ is $\{\texttt{ID}, \texttt{CURRENCY}, \texttt{TITLE}\}$, which includes \texttt{CURRENCY} to specify the nature of the target values, ensuring a well-rounded contextual representation.
    \end{example}

\cref{alg:retrieval} depicts the retrieval method.
Parameter $K$ is  the desired size of \colcontext. 
If a table $T$ has columns less than $K$, we simply use all available columns (excluding the target) as $\candpool$.
Otherwise, we first compute the column embeddings for all columns in $T$ using the text encoder $\mathcal{E}$ (Lines 1-2).
We then initialize the column context $\candpool$ as empty and select the first column $c'$ to be added into $\candpool$ as the most similar column to the target (Lines 4-5). 
Next, we iteratively select the next column $c'$ from $T \setminus \candpool$ based on its marginal relevance score $\mr(c,\target,\candpool)$, until the size of $\candpool$ reaches $K$ (Lines 6-8). 
Specifically, the selected $c'$ maximizes the marginal relevance score (Line 7) and is added to $\candpool$ (Line 8).
In experiments, we have varied $K$ to study its impact in \cref{fig:exp_size}; the performance increases first and then remains stable after a certain $K$ value, indicating that a moderate number of contextual columns is sufficient to provide a well-rounded context for the annotation tasks.

\subsection{Context-Aware Encoding}
\label{sec::encoding}
After retrieving  $\candpool$ for a target $\target$ in a table $T$, we need to effectively leverage it   for annotation.
For different targets, their context columns may vary, and the model should be trained to prioritize the target while recognizing the supporting role of its context columns.
To achieve this, we   propose a \textit{context-aware encoder} that incorporates column role embeddings to explicitly differentiate the target from its context columns during the embedding process. Additionally, we introduce \textit{target-context pair training} to train the model on individual target-context pairs, ensuring that each target is paired with its specific context rather than entire tables in a single pass.

\begin{figure}[t]
    \centering
      \includegraphics[width=0.39\textwidth]{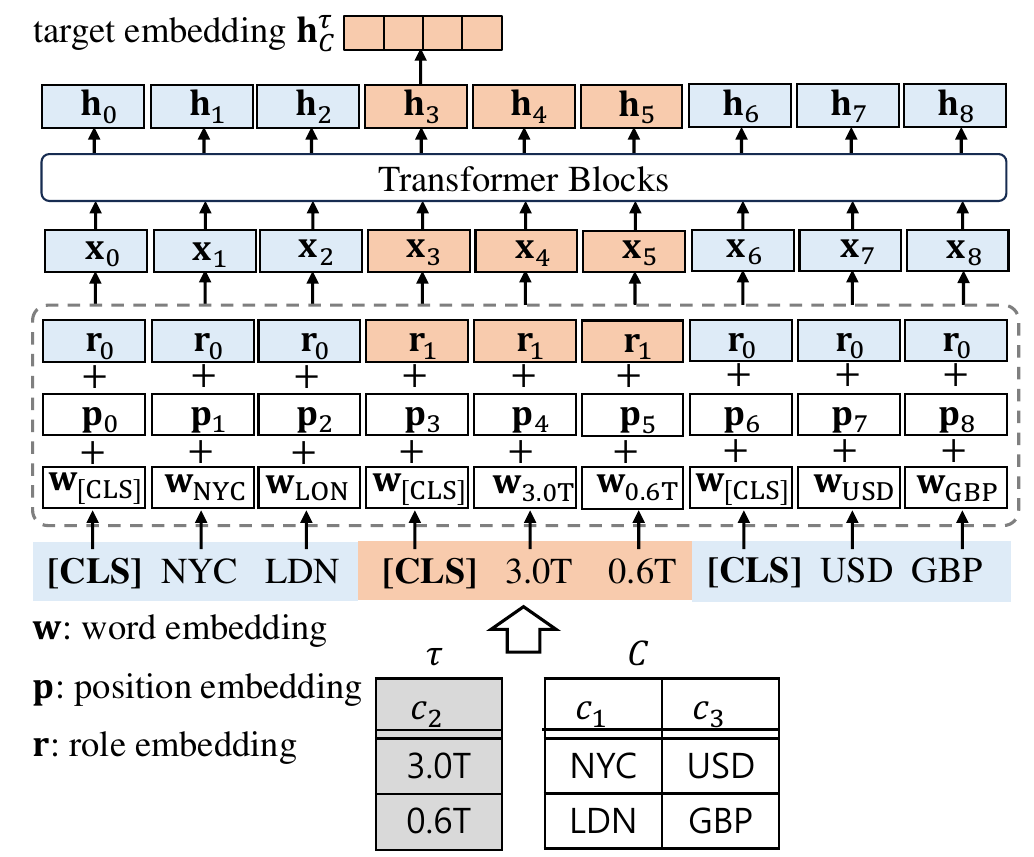}
      \vspace{1mm}
    \caption{Illustration of Context-Aware Column Embedding.}
    \label{fig:encoding}
    \vspace{1mm} 
\end{figure}

\header 
\textbf{Context-Aware Encoder with Role Embeddings.} 
Given a target $\target$ and its \colcontext $\candpool$ in a table $T$, we first serialize the columns in $\candpool$ into a token sequence by preserving their original order in the table and concatenating their cell values column by column, separated by a special token (\texttt{[CLS]} in BERT~\cite{bert}) used in language models.  For instance, consider the table in \cref{fig:encoding}: the serialization process produces the sequence ``\texttt{[CLS]} NYC LDN \texttt{[CLS]} 3.0T 0.6T \texttt{[CLS]} USD GBP'', where each column is delineated by a \texttt{[CLS]} token.
Then for each token $w_j$ in the sequence, the language model maps it to an embedding by aggregating a pretrained word embedding $\mathbf{w}_{j}$ and a position embedding $\mathbf{p}_{j}$, which encodes the position of the token in the sequence.

To explicitly distinguish the target from its context columns, we introduce the third embedding component for a token, \textit{column role embedding}, in the context-aware encoder. 
For each token $w_j$ in the serialized sequence, we assign a binary role indicator $r_j \in \{0, 1\}$, where $r_j = 1$ if the token belongs to the target and $r_j = 0$ otherwise. 
This role indicator is mapped to a role embedding vector $\mathbf{r}_j \in \mathbb{R}^d$ using a trainable lookup table $\mathbf{E}_{\mathrm{role}} \in \mathbb{R}^{2 \times d}$, where $d$ is the embedding dimension, as follows.
\begin{equation}
    \small
    \mathbf{r}_j = \mathbf{E}_{\mathrm{role}}[r_j],
    \end{equation}
where $\mathbf{r}_j  \in \mathbb{R}^d$  is the role embedding of token $w_j$. 

Note that $\mathbf{E}_{\mathrm{role}}$ contains two embeddings: one for tokens belonging to the target and another for tokens from context columns. These embeddings are jointly optimized with the other model parameters during training, enabling the model to effectively differentiate structural roles and prioritize the target.

Then, as shown in Figure \ref{fig:encoding}, a token $w_j$ is embedded into $\mathbf{x}_{j}$ that combines three components: the word embedding $\mathbf{w}_{j}$, position embedding $\mathbf{p}_{j}$, and role embedding $\mathbf{r}_{j}$ in \cref{eq:token}.
\begin{equation}\label{eq:token}
    \small
\mathbf{x}_{j} = \mathbf{w}_{j} + \mathbf{p}_{j} + \mathbf{r}_{j}.  
\end{equation}

Then the token embeddings $\mathbf{x}_{j}$ are fed into a language model, primarily composed of transformer layers with self-attention mechanisms~\cite{transformer}, 
to produce the contextualized target embedding $\mathbf{h}^{\target}_{\candpool}$ for target $\target$. Specifically, $\mathbf{h}^{\target}_{\candpool}$ is extracted from the output corresponding to the \texttt{[CLS]} token of the target in CTA, regarding \texttt{[CLS]} as a representative token for the column, as shown in Figure \ref{fig:encoding}. 
In CPA, with a pair of two columns as the target, we concatenate the embeddings of the   columns to form the target embedding. 
In our design, $\mathbf{h}^{\target}_{\candpool}$ captures the semantics of the target, while also incorporating contextual information from the columns in $\candpool$, considering their distinct roles.
Then a linear layer classifier $\psi$ is applied to   $\mathbf{h}^{\target}_{\candpool}$ to predict the semantic type type or relation of the target,  
\begin{equation}\label{eq:predict} 
    \small
    f(\target, \candpool) = \psi(\mathbf{h}^{\target}_{\candpool}), 
\end{equation} 
where the prediction module $f(\target, \candpool)$ denotes the predictions for the target $\target$ based on the given context $\candpool$.

\header
\textbf{Target-Context Pair Training.} 
Prior works~\cite{doduo, starmie, watchog} typically use the entire table $T$ as the training instance for different targets, resulting in all targets sharing the same table context. This approach may include columns that are irrelevant or uninformative for specific targets, potentially degrading model performance. 
In contrast, our design allows different targets in the same table $T$ to have distinct \colcontexts $\candpool$, tailored to each target. 
After obtaining the \colcontext $\candpool$ for a target $\target$ in a table $T$, we pair the target $\target$ with its unique context columns $\candpool$ and the corresponding ground-truth label $\gt$, forming a training instance $(\target, \candpool, \gt)$. 
This reformulates the training data into a set of target-context pairs, enabling the model to focus on target-specific contexts. 
Since the retrieval method in \cref{sec:retrieval} is unsupervised, this process is efficient and does not require additional supervision. 
Formally, for each labeled target in the original training set $\datasett$, we construct the target-context pair training set $\datasetp$ as follows:
\begin{equation}\label{eq:dataset}
\small
\datasetp = \{
(\target, \candpool, \gt) \mid \target \in \datasett
\},
\end{equation}
where $\gt$ is the ground truth of $\target$.

Our method \model is then trained using this target-context pair training set, optimizing the following objective: 
\begin{equation}\label{eq:cta_loss}
    \small
    \mathcal{L}(\theta) = \frac{1}{|\datasetp|} \sum_{(\target, \candpool, \gt)\in \datasetp } \ell(f(\target,\candpool), \gt)  
\end{equation}
where $f(\target,\candpool)$ is predication for target $\target$, and $\ell$ is the cross-entrpoy loss, and $\theta$ represents the trainable parameters.

\section{The REVEAL+ Method}
\label{sec:verification}  
Given a target $\target$ in a table $T$, we efficiently obtain its column context $\candpool$ in \cref{sec:retrieval} using unsupervised information, which is independent of specific CTA/CPA tasks. 
However, the retrieved context columns in $\candpool$ are \textit{not yet verified} for their effectiveness in the specific CTA/CPA tasks. 
In this section, we propose \modelplus, which incorporates a \textit{verification method} to refine $\candpool$ by directly evaluating the effectiveness of its columns for the target $\target$ for the annotation tasks. This refinement produces a \textit{verified column context} $\vpool$, which improves the annotation performance of \modelplus compared to \model. 
We first formulate the column verification problem as a classification task and present the verification model design (\cref{sec::verification_model}). To further enhance efficiency, we propose a top-down verification inference technique to greedily identify $\vpool$ from $\candpool$ without exhaustively evaluating all subsets (\cref{sec::inference}).

\subsection{Verification Model Formulation and Design}\label{sec::verification_model}
During inference, given a table $T$, we aim to identify a verified column context $\vpool \subseteq \candpool$ that is most effective for the target $\target$. 
A naive approach is to exhaustively evaluate all possible subsets of $\candpool$ using the trained prediction module component $\tmodel$ (as shown in \cref{fig:framework}(a)) and select the subset with the highest prediction confidence as $\vpool$. 
However, prior research has demonstrated that model confidence is often poorly calibrated and does not reliably indicate prediction correctness~\cite{DBLP:conf/icml/GuoPSW17}, result in suboptimal or misleading decisions, as validated by experiments in \cref{sec::sensitivity}. Moreover, exhausting all subsets is computationally expensive.

On the other hand, we leverage the training and validation datasets with ground truth to train a lightweight verification model $\vmodel$, which 
produces a quality score $\vmodel(\target,\vcan)$  to evaluate the quality of a subset $\vcan \subseteq \candpool$ as a verified column context for the target $\target$. Below, we formulate column context verification as a classification task. 
The formal column context verification problem is  to find $\vpool$ that maximizes the quality score $\vmodel(\target,\vcan)$, as follows:

\begin{definition}[Column Context Verification] \label{def:verification} 
Given a table $T$ with a target $\target$ and its column context $\candpool$, let $\mathbb{S} = \{ \vcan \mid \vcan \subseteq \candpool \}$. Column context verification aims to identify the subset $\vpool \in \mathbb{S}$ that maximizes the quality score, i.e., 
\begin{equation}
\vpool = \underset{\vcan \in \mathbb{S}}{\operatorname{argmax}} \ \vmodel(\target, \vcan),
\end{equation} 
where $\vmodel$ evaluates the effectiveness of $\vcan$ for annotating $\target$.

    \end{definition}

A key challenge in implementing the verification model $\vmodel$ is the lack of labeled training data. There are no predefined labels indicating whether a subset of columns $\vcan$ is effective for annotating a target $\target$. Manually defining rules for labeling may not align with actual annotation performance and fails to account for the target-specific nature of context quality. Different targets within the same table often require distinct context columns, making the labeling process inherently complex and context-dependent.
To address this, we propose to construct labeled data for $\vmodel$ using the training and validation datasets, as these datasets already contain ground-truth labels. 
This approach ensures that the verification model $\vmodel$ is aligned with the trained prediction module $\tmodel$ in \cref{fig:framework}(a), as $\vmodel$ is trained using the predictions of $\tmodel$ on these datasets.

\header \textbf{Labeled Data for Verification Model.} 
Given a training or validation sample with a target $\target$ in a table $T$ and its retrieved column context $\candpool$, we determine whether a subset $\vcan \subseteq \candpool$ enables the trained prediction module $\tmodel$ to make a correct prediction for $\target$. Specifically, for each subset $\vcan$, we obtain the prediction $\tmodel(\target,\vcan)$ via \cref{eq:predict}, and compare it with the ground-truth label $\gt$. This binary outcome (correct or incorrect prediction) serves as the label $\vlabel$ for training our verification model. Formally, as defined in \cref{eq:label}, if the prediction $\tmodel(\target, \vcan)$ matches the ground truth $\gt$, we label $\vlabel=1$ (positive, i.e., high quality); otherwise, we label $\vlabel=0$ (negative, i.e., low quality). 
\begin{equation}\label{eq:label}
    \vlabel =
    \begin{cases}
        1, & \text{if } \tmodel(\target, \vcan) = \gt,  \\
        0, & \text{otherwise.} 
   \end{cases}
\end{equation}

To create the labeled data $\datasetvt$ for the verification model, for each training and validation table $T$ with target $\target$, ground truth $\gt$, and column context $\candpool$, we evaluate every subset $\vcan \subseteq \candpool$. Using $\tmodel$, we compute the prediction $\tmodel(\target,\vcan)$ and assign the label $\vlabel$ for $\vcan$ based on \cref{eq:label}, resulting in a labeled sample $(\target, \vcan, \vlabel)$. 

Both training and validation datasets are used to construct $\datasetvt$ for the following reasons. The prediction module $\tmodel$, trained on the training dataset, tends to produce mostly positive labels ($y^\target_{\vcan} = 1$), which may lead to insufficient negative samples. In contrast, the validation dataset, unseen during $\tmodel$ training, provides more negative examples ($y^\target_{\vcan} = 0$), ensuring a balanced dataset for training $\vmodel$.

The construction of $\datasetvt$ occurs after $\tmodel$ is trained, ensuring no interference between their training processes. During inference, the trained $\vmodel$ is used to identify the verified column context $\vpool$, which enhances the predictions of the trained $\tmodel$, as shown in \cref{fig:framework}(b).

\header \textbf{Verification Model Design.}
The architecture of the verification model $\vmodel$ is designed to predict the label $\vlabel$ (0 or 1) for a given target $\target$ in table $T$ and a subset of columns $\vcan$. A simple implementation could treat $\vmodel$ as a binary classifier, outputting a single logit transformed by a sigmoid function to estimate $\vmodel(\target, \vcan)$, the probability that $\tmodel$ correctly predicts $\target$'s label given context $\vcan$. However, as noted in \cite{DBLP:conf/iclr/HendrycksG17}, such an approach may lead to overconfident positive predictions, as it does not explicitly model the likelihood of negative outcomes.

To address this, we implement $\vmodel$ as a lightweight multilayer perceptron (MLP) with an output layer producing two logits: one for the positive class and one for the negative class. The softmax probability of the positive class is used as the \qscore. This design ensures that when the verification model is uncertain about a context, the logits for both classes are close, resulting in a softmax score near 0.5. Conversely, when the model is confident, the logits are skewed, yielding a score closer to 0 or 1.
We use the context-aware encoding technique from \cref{sec::encoding} to compute the embedding $\mathbf{h}^{\target}_{\vcan}$ for the target $\target$ given the subset $\vcan$. The MLP in the verification model then processes $\mathbf{h}^{\target}_{\vcan}$ and outputs a two-dimensional logit vector $\mathbf{z}_{\vcan}^{\target} \in \mathbb{R}^2$, corresponding to the logits for the negative and positive classes:
\begin{equation}\label{eq:verifier_logits}
    \small
    \mathbf{z}_{\vcan}^{\target} = \operatorname{MLP}(\mathbf{h}^{\target}_{\vcan}).
    \end{equation}

The predicted \qscore is then computed as the softmax probability of the positive class:
\begin{equation}\label{eq:softmax}
    \small
\vmodel(\target, \vcan) = \frac{e^{\mathbf{z}_{\vcan,1}^{\target}}}{\sum_{j=0}^{1}e^{\mathbf{z}_{\vcan,j}^{\target}}}.
\end{equation}

The loss function $\mathcal{L}_{\mathrm{v}}$ for training the verification model  is 
\begin{equation}\label{eq:ce}
\resizebox{1\linewidth}{!}{$
\mathcal{L}_{\mathrm{v}}(\theta_{\vmodel}) = 
- \sum\limits_{(\target, \vcan, y_{\vcan}^{\target}) \in \datasetvt}
\left[
(1 - y_{\vcan}^{\target}) \log(1 - \vmodel(\target, \vcan)) +
y_{\vcan}^{\target} \log \vmodel(\target, \vcan)
\right],
$}
\end{equation}
where $\theta_{\vmodel}$ represents the trainable parameters.

\subsection{Top-Down Verification Inference }\label{sec::inference}
After training the verification model $\vmodel$ in \cref{sec::verification_model}, we use it during inference to identify the verified column context $\vpool$ from $\candpool$ for the target $\target$. However, evaluating all $2^{|\candpool|}$ possible subsets of $\candpool$ is computationally prohibitive. Instead, we propose a top-down inference method to efficiently search for $\vpool$ in a greedy manner.

The method starts with the full $\candpool$ and iteratively removes columns that are less informative to the target $\target$. This approach leverages the fact that the retrieved context columns in $\candpool$ from \cref{sec:retrieval} are already high-quality, with only a few potentially noisy or irrelevant ones. Larger contexts generally provide richer semantic information, offering a more comprehensive understanding of the target. However, excessively small subsets may amplify the influence of individual misleading columns, degrading predictions. The top-down verificaiton process makes greedy decisions and incorporates early stopping to balance efficiency and effectiveness.

Specifically, let $\vpool^{(t)}$ denote the verified column context obtained in the $t$-th iteration.
The top-down inference starts from $\vpool^{(t=0)}=\candpool$ of size $|\candpool|$ for the target $\target$ in a table $T$.
In the next $(t+1)$ iteration, we generate all subsets of $\vpool^{(t)}$ by removing a single column from it, i.e., $\vcan \subset \vpool^{(t)}$ with $|\vcan|=|\vpool^{(t)}|-1$.
For all these subsets, we compute their quality scores $\vmodel(\target, \vcan)$  by \cref{eq:verifier_logits,eq:softmax}.
We then select the subset $\vcan$ with the highest quality score as the new verified column context $\vpool^{(t+1)}$, to be used in the next iteration.
The computation is formally described as:
\begin{equation} 
    {\vpool}^{(t+1)} = \underset{\vcan \subset {\vpool}^{(t)},\, |\vcan|=|\vpool^{(t)}|-1}{\operatorname{argmax}}\, \vmodel(\target, \vcan),
\end{equation}

This greedy refinement process continues until the size of $\vpool^{(t)}$ reaches 1 or the following early stop condition is met: 
If the new verified column context $\vpool^{(t+1)}$ has a lower quality score than the previous one $\vpool^{(t)}$, meaning that unlikely the new $\vpool^{(t+1)}$ can help to make better predictions for $\target$, and the subsequent iterations may not yield better results.
In this case, we stop the refinement process and keep the current  $\vpool^{(t)}$ as the final verified column context $\vpool$ for the input target $\target$.
Formally, we check the following early-stop condition:
\begin{equation}  
    \vmodel(\target, \vpool^{(t+1)}) < \vmodel(\target, \vpool^{(t)}).
\end{equation}

This top-down inference technique reduces the search space from $2^{|\candpool|}$ to at most $|\candpool|^2$ (in the worst case). In practice, it is much smaller due to early stopping.

\SetKwInOut{Parameter}{Variables}

\begin{algorithm}[!t]
\caption{Top-Down Inference for Verified Column Context}
\label{alg:topdown_inference}
\small
\KwIn{target $\target$, column context $\candpool$, trained verification model $\vmodel$}
\KwOut{Verified column context $\vpool$}

$t \gets 0$; \\
$\vpool^{(t)} \gets \candpool$; \\
Get $\vmodel(\target, \vpool^{(t)})$ by 
\cref{eq:verifier_logits,eq:softmax}; \\

\While{$|\vpool^{(t)}| > 1$}{
    $\mathbb{S} \gets \{\vcan \subset \vpool^{(t)} \mid |\vcan| = |\vpool^{(t)}| - 1\}$; \\
    $\vpool^{(t+1)} \gets \emptyset$; \\
    $\vmodel(\target, \vpool^{(t+1)}) \gets -\infty$; \\
    \ForEach{$\vcan \in \mathbb{S}$}{
        Get $\vmodel(\target, \vcan)$ by \cref{eq:verifier_logits,eq:softmax}; \\
        \If{$\vmodel(\target, \vcan) > \vmodel(\target, \vpool^{(t+1)})$}{
            $\vpool^{(t+1)} \gets \vcan$; \tcp{Update best subset}
            $\vmodel(\target, \vpool^{(t+1)}) \gets \vmodel(\target, \vcan)$; \tcp{Update best score}
        }
    }
    \If{$\vmodel(\target, \vpool^{(t+1)}) < \vmodel(\target, \vpool^{(t)})$}{
        \textbf{break}; \tcp{Early stopping}
    }
    $t \gets t + 1$;  
}
$\vpool \gets \vpool^{(t)}$;\\
\Return $\vpool$;
\end{algorithm}

\header\textbf{Algorithm.}
\cref{alg:topdown_inference} summarizes the top-down inference process. 
In Lines 1-3, we initialize the verified column context $\vpool^{(0)}$ as $\candpool$ and compute its quality score using the verification model $\vmodel$ with \cref{eq:verifier_logits,eq:softmax}.
In Lines 4-15, we iteratively refine the verified column context $\vpool^{(t)}$ by generating all subsets of size $|\vpool^{(t)}|-1$ and selecting the one with the highest quality score.
The while loop terminates when the size of $\vpool^{(t)}$ reaches 1 or the early stop condition is met at Lines 13-14.
In the $t$-th iteration, we generate all subsets of size $|\vpool^{(t)}|-1$ (Line 5), and initialize the best subset $\vpool^{(t+1)}$ as empty and its quality score as $-\infty$ (Line 6-7).
Then we iterate over all subsets $\vcan$  in $\mathbb{S}$ (Line 8) and compute their quality scores using the verification model $\vmodel$ (Line 9). 
If the current subset $\vcan$ has higher $\vmodel(\target, \vcan)$ than the best subset $\vpool^{(t+1)}$ so far, we update the best subset $\vpool^{(t+1)}$ and its quality score to be the current subset $\vcan$ and its quality score $\vmodel(\target, \vcan)$ (Line 10-12).
Lines 13-14 check the early stop condition. If the new verified column context $\vpool^{(t+1)}$ has a lower quality score than the previous one $\vpool^{(t)}$, we stop the  process. 
Otherwise, we update the iteration index $t$ and continue to the next iteration (Line 15). 
Finally, we return the verified column context $\vpool$ as the final output (Lines 16-17).

\subsection{Training Process}\label{sec::training_process}

We summarize the training process of \modelplus in \cref{alg:modelplus_training}.
Given a training set $\datasett$, we obtain the column contexts $\candpool$ for the targets using the retrieval algorithm in \cref{sec:retrieval} (Line 1). 
Then we  obtain the target-context pair training set $\datasetp$ using \cref{eq:dataset} (Line 2). As introduced in \cref{sec::encoding}, this dataset includes pairs of target columns (or column pairs for CPA) and their retrieved context columns.
Then, we train the prediction module $\tmodel$ using  $\datasetp$ and the loss function $\mathcal{L}(\theta)$ defined in ~\cref{eq:cta_loss} (Line 3). Once the prediction module $\tmodel$ is trained, we construct the verification training set $\datasetvt$ from $\datasetp$ as described in ~\cref{eq:label} (Line 4). This set consists of labeled context subsets based on whether they enable correct predictions by the trained prediction module.
Finally, in Line 5, we train the verification model $\vmodel$ using $\datasetvt$ with the binary cross-entropy loss $\mathcal{L}_{\mathrm{v}}(\theta_{\vmodel})$  defined in ~\cref{eq:ce}. The procedure concludes by returning the trained $\tmodel$ and $\vmodel$ (Line 6).
When Lines 4 and 5 are skipped, it is the training process of \model, which only involves context retrieval and training the prediction module $\tmodel$.

The inference processes of \model and \modelplus have already been illustrated in Figure \ref{fig:framework}.

\begin{algorithm}[!t]
\caption{\modelplus Training Procedure}
\label{alg:modelplus_training}
\small
\KwIn{training set $\datasett$, initialized prediction module $\tmodel$ and verification model $\vmodel$}
\KwOut{trained prediction model $\tmodel$, verification model $\vmodel$}

Obtain column contexts $\candpool$ for training targets  in $\datasett$ by Algorithm~\ref{alg:retrieval};

Construct $\datasetp$ from $\datasett$ by ~\cref{eq:dataset}; 

Train $\tmodel$ on $\datasetp$ using $\mathcal{L}(\theta)$ in~\cref{eq:cta_loss}; 

Construct $\datasetvt$ from $\datasetp$ by ~\cref{eq:label};

Train $\vmodel$ on $\datasetvt$ using $\mathcal{L}_{\mathrm{v}}(\theta_{\vmodel})$ in~\cref{eq:ce}; 

\Return trained $\tmodel$, $\vmodel$;
\end{algorithm}

\section{Experiments}
\label{sec:exp}
 
We conduct extensive experiments to evaluate our proposed methods, \model and \modelplus, on real-world datasets for both CTA and CPA tasks. The implementation is publicly available at: \url{https://github.com/TommyDzh/REVEAL}.

\subsection{Experiment Setup}\label{sec::exp_setup}

\header\textbf{Datasets.} 
We use six benchmark datasets with real-world tables, four for CTA (\gitdb, \gitsc, \sotabcta, \wikicta) and two for CPA (\sotabcpa, \wikicpa). 
\Cref{tab:dataset} summarizes the dataset statistics. 
\gitdb and \gitsc are part of the SemTab 2022 benchmark~\cite{semtab}, sourced form GitTables~\cite{gittables}, and are both used for the CTA task. \gitdb columns are annotated with DBpedia properties, while \gitsc uses Schema.org properties. The tables in the two datasets are wide. For example, the average of  columns per table in \gitdb is 12.1, with some tables having up to 193 columns.  
The \sotabcta and \sotabcpa datasets are part of the WDC Schema.org Table Annotation Benchmark (SOTAB)~\cite{sotab}, which targets the CTA and CPA tasks using 91 Schema.org types and 176 Schema.org relations, respectively. The number of columns in the tables in these datasets can be up to 31. 
The \wikicta and \wikicpa datasets are based on the WikiTables corpus from Wikipedia and introduced by TURL~\cite{turl}. They are used for CTA and CPA, respectively, with 255 DBpedia types and 121 DBpedia relations. Note that in these datasets, we uses all columns from the original tables. 
The average number of columns per table in \wikicta and \wikicpa is not large, but the tables can still be wide with the average number of columns being 5.9 and 5.5, respectively. 

\begin{table}[!t]
  \centering 
  \caption{Dataset statistics.}  
  \label{tab:dataset}
  \vspace{-4mm}
  \resizebox{1\linewidth}{!}{
    \setlength{\tabcolsep}{2pt}
      \renewcommand{\arraystretch}{0.92}
    \begin{tabular}{lccccl}
      \toprule
      Benchmark & \# Tables & \# Types & Total \# Cols & \# Labeled Cols & \begin{tabular}[c]{@{}c@{}}Min/Max/Avg\\ Cols per Table\end{tabular} \\
      \midrule
      \gitdb      & 3,737     & 101  & 45,304     & 5,433     & 1 / 193 / 12.1 \\
      \gitsc      & 2,853     & 53   & 34,148     & 3,863     & 1 / 150 / 12.0 \\
      \sotabcta   & 24,275    & 91   & 195,543    & 64,884    & 3 / 30 / 8.1   \\
      \sotabcpa   & 20,686    & 176  & 196,831    & 74,216    & 3 / 31 / 9.5   \\
      \wikicta    & 406,706   & 255  & 2,393,027  & 654,670   & 1 / 99 / 5.9   \\
      \wikicpa    & 55,970    & 121  & 306,265    & 62,954    & 2 / 38 / 5.5   \\
      \bottomrule
    \end{tabular}
  }
\end{table}

\begin{table*}[t]
  \centering
  \caption{Overall performance by \micro(\%) and \macro(\%) (mean ± std).}
  \vspace{-3mm}
    \resizebox{1\textwidth}{!}{
   \setlength{\tabcolsep}{3pt}
    \begin{tabular}{ccccccccccccc|c}
    \toprule
    \multirow{2}[4]{*}{Method} & \multicolumn{2}{c}{\gitdb} & \multicolumn{2}{c}{\gitsc} & \multicolumn{2}{c}{\sotabcta} & \multicolumn{2}{c}{\sotabcpa} & \multicolumn{2}{c}{\wikicta} & \multicolumn{2}{c|}{\wikicpa} & \multirow{2}[4]{*}{Avg. Rank} \\
\cmidrule{2-13}    & \micro & \macro & \micro & \macro & \micro & \macro & \micro & \macro & \micro & \macro & \micro & \macro &  \\
    \midrule
    \turl & 48.20 ± 0.98 & 19.56 ± 2.45 & 58.15 ± 0.98 & 26.26 ± 4.70 & 80.13 ± 0.25 & 77.34 ± 0.98 & 66.11 ± 0.31 & 60.52 ± 0.27 & 90.53 ± 0.07 & 66.40 ± 0.23 & 89.60 ± 0.03 & 80.49 ± 0.36 & {7.42}  \\
    \sato & 46.05 ± 0.87 & 23.50 ± 1.01 & 55.83 ± 0.34 & 30.07 ± 0.94 & 71.74 ± 0.25 & 70.01 ± 0.27 & 54.06 ± 0.15 & 46.59 ± 0.24 & 61.56 ± 0.84 & 32.05 ± 0.72 & 77.09 ± 0.15 & 49.70 ± 0.26 & { 8.33}  \\
    \doduo & 44.77 ± 4.36 & 21.63 ± 3.26 & 52.36 ± 2.94 & 25.09 ± 5.01 & 81.32 ± 0.00 & 78.51 ± 0.00 & 78.85 ± 0.49 & 74.95 ± 0.49 & 92.19 ± 1.01 & 74.10 ± 0.86 & 91.60 ± 0.23 & 83.81 ± 0.15 & {6.58}  \\
    \starmie & 54.43 ± 0.93 & 30.71 ± 3.58 & 64.87 ± 2.47 & 37.04 ± 4.33 & \uuline{87.95 ± 0.27} & \uuline{86.84 ± 0.08} & \uuline{78.92 ± 0.13} & \uuline{75.69 ± 0.31} & \uuline{92.91 ± 0.08} & \uuline{76.45 ± 0.44} & \uuline{92.51 ± 0.36} & \uuline{85.69 ± 0.47} & {\uuline{3.50}} \\
   \reca & 53.83 ± 1.67 & 25.70 ± 5.49 & 64.52 ± 1.69 & 35.10 ± 3.81 & 68.52 ± 0.35 & 68.21 ± 0.71 & 54.31 ± 0.99 & 47.56 ± 0.95 & 91.27 ± 0.03 & 73.36 ± 0.82 & 88.90 ± 0.94 & 78.39 ± 1.09 &{ 7.17 } \\
  \watchog & 53.96 ± 0.78 & 28.89 ± 3.34 & 65.24 ± 1.90 & 36.06 ± 3.39 & 86.23 ± 0.22 & 84.19 ± 0.06 & 76.52 ± 0.11 & 72.60 ± 0.25 & 92.25 ± 0.06 & 73.77 ± 0.35 & 92.28 ± 0.00 & 85.27 ± 0.25 & {4.75}  \\
    \midrule
    {\qwen (0-shot)} & 53.51 ± 0.60 & 27.75 ± 1.81 & \uuline{66.80 ± 1.82} & 40.35 ± 2.43 & 48.26 ± 0.00 & 44.36 ± 0.00 & 52.68 ± 0.00 & 44.01 ± 0.00 & 31.70 ± 0.00 & 12.28 ± 0.00 & 27.63 ± 0.00 & 18.13 ± 0.00 & { 8.67 } \\
    \qwen (5-shot) & \uuline{57.48 ± 1.20} & \uuline{33.66 ± 1.57} & 63.21 ± 1.24 & \underline{44.81 ± 2.60} & 52.61 ± 0.00 & 48.82 ± 0.00 & 53.48 ± 0.00 & 45.41 ± 0.00 & 32.21 ± 0.00 & 13.01 ± 0.00 & 28.18 ± 0.00 & 18.35 ± 0.00 & 7.58 \\
   \tllama & 11.03 ± 0.74 & 9.70 ± 1.53 & 12.85 ± 0.91 & 11.71 ± 1.62 & 23.35 ± 0.00 & 20.59 ± 0.00 & 16.96 ± 0.00 & 13.81 ± 0.00 & 91.85 ± 0.00 & 76.04 ± 0.00 & 92.12 ± 0.00 & 85.67 ± 0.00 & {8.92}\\
    \midrule
    \model & \underline{59.79 ± 0.98} & \underline{36.40 ± 3.17} & \underline{69.26 ± 0.93} & \uuline{41.98 ± 1.96} & \underline{88.35 ± 0.00} & \underline{87.60 ± 0.00} & \underline{80.45 ± 0.25} & \underline{77.81 ± 0.24} & \textbf{93.14 ± 0.51} & \underline{77.81 ± 0.43} & \underline{92.76 ± 0.11} & \underline{86.40 ± 0.07} & {\uline{2.00}}  \\
   \modelplus & \textbf{61.53 ± 2.27} & \textbf{38.30 ± 1.16} & \textbf{70.90 ± 2.18} & \textbf{45.82 ± 1.16} & \textbf{88.74 ± 0.22} & \textbf{88.10 ± 0.06} & \textbf{80.81 ± 0.07} & \textbf{78.15 ± 0.15} & \underline{93.00 ± 0.04} & \textbf{78.07 ± 0.22} & \textbf{92.80 ± 0.18} & \textbf{86.81 ± 0.23} & {\textbf{1.08}}  \\
    \bottomrule
    \end{tabular}%
    }
  \label{tab:overall}%
      \vspace{-2mm}
\end{table*}%

\begin{figure*}[t]
  \centering
    \includegraphics[width=1\textwidth]{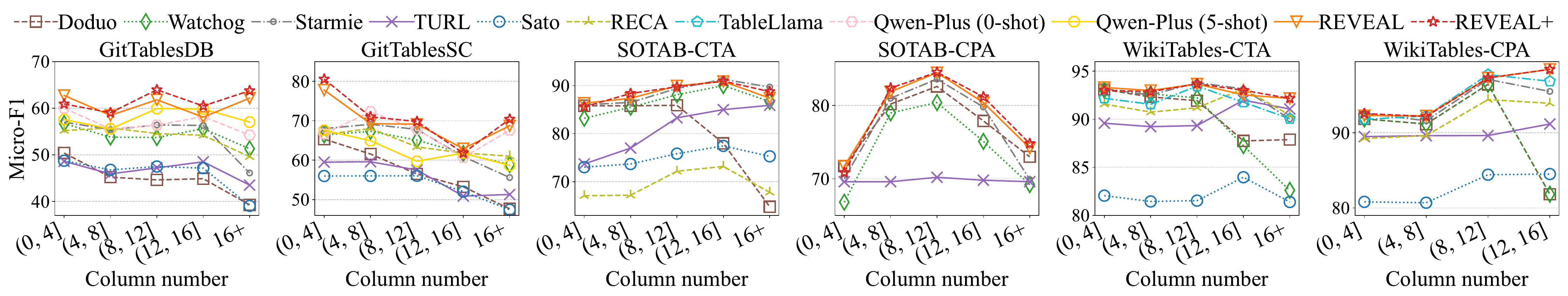}
  \caption{Performance on tables with a small to large number of columns.}
  \label{fig:column_num}
      \vspace{-2mm}
\end{figure*}

\header\textbf{Baselines.}  
We compare \model and \modelplus against strong baselines, including six state-of-the-art models specifically designed for annotation tasks and two large language models (LLMs).
\vspace{-\topsep}
\begin{itemize}[leftmargin=*]
    \item \textbf{\turl}~\cite{turl} adapts the transformer architecture to better capture table structure. It is pretrained on a large-scale table corpus.
    
    \item \textbf{\sato}~\cite{sato} uses table-level features and models pairwise dependencies between neighboring columns for column type predictions.
    
    \item \textbf{\doduo}~\cite{doduo}  introduces a serialization strategy for tables, enabling language models to effectively encode
    columns for annotation tasks. It is fine-tuned via multi-task training.
    
    \item \textbf{\reca}~\cite{reca}  utilizes named entity-based schema alignment to consider inter-table information.

    \item \textbf{\starmie}~\cite{starmie} develops a contrastive multi-column pretraining to enhance embeddings by incorporating  semantics  within tables.
    
    \item \textbf{\watchog}~\cite{watchog} leverages contrastive learning on unlabeled table corpora to learn  representations for table understanding tasks.
    
    \item \textbf{\tllama}~\cite{tablellama} is an open-source generalist model for a various table-based tasks. It fine-tunes LLMs on an extensive corpus of  tables and associated tasks.
    
    \item \textbf{\qwen (0-shot / 5-shot)}~\cite{DBLP:journals/corr/abs-2407-10671} is an LLM with strong capabilities on NLP and table-based tasks. We evaluate both 0-shot and 5-shot in-context learning variants, following the setup in~\cite{korini2024column,korini2023column}.
\end{itemize}

\header\textbf{Evaluation Metrics.}  
For effectiveness, we adopt the standard evaluation metrics used in prior work~\cite{turl, gittables, sotab} for CTA and CPA tasks, including \micro and \macro, since the tasks are multi-class classification problems.  
Following prior work, we evaluate performance using ground-truth labels, without considering semantic hierarchies or specificity. 
A prediction is correct if it matches the ground truth.
\micro  is the weighted average of F1 scores of classes, where each class contributes proportionally to its number of samples. It reflects overall performance but is biased toward frequent classes.
\macro is the unweighted average of F1 scores across all classes, treating each class equally. It is more sensitive to performance on rare or underrepresented classes.
We perform 5-fold cross-validation on \gitdb and \gitsc following~\cite{watchog}. For the remaining datasets, we train and evaluate each model five times with different random seeds, except the LLM baselines that directly use task-specific prompts for predictions~\cite{tablellama} as explained below. We report the mean and standard deviation of results over five runs.

\header\textbf{Implementations.} 
For baselines, we obtain their publicly available codebases and adopt them for the CTA and CPA tasks. 
We adopt BERT as the base language model for \doduo, \starmie, \reca, \turl and \watchog to ensure a fair comparison. 
For LLM-based methods \tllama and \qwen, we utilize task-specific prompts for CTA and CPA, following the prompt templates provided in \cite{tablellama}, to generate predictions for column types and relations. We run the offically released \tllama and 
access  Qwen-Plus API for   0-shot and 5-shot in-context predictions. For 5-shot, demonstrations are randomly sampled from the training set for each test instance, following~\cite{korini2024column,korini2023column}.
We implement our methods \model and \modelplus in Python using PyTorch and the Transformer library~\cite{DBLP:conf/emnlp/WolfDSCDMCRLFDS20}. The context-aware encoder is built on BERT as well. We use the Adam optimizer to train.
The learning rate is set to 5e-5, the desired size of column context $K$ is fixed at 8 across all datasets, and the maximum input sequence length is set to 256 tokens. 
For long columns, we follow prior work~\cite{doduo,watchog} by allocating tokens equally across columns and truncating row-wise, using the earliest rows that fit within the token limit.
We select the best model checkpoint based on the highest Micro-F1 or Macro-F1 score on validation data. For the verification model in \modelplus, it is a three-layer MLP trained after the annotation model. We use a batch size of 64 for \gitdb and \gitsc, and 512 for the other datasets. The learning rate is tuned from $\{5\mathrm{e}{-5}, 1\mathrm{e}{-4}, 5\mathrm{e}{-4}\}$. 
All experiments are conducted on a Linux server with an Intel Xeon Gold 6226R CPU with 2.90GHz and an NVIDIA RTX 3090 GPU.

\subsection{Overall Results}\label{sec::overall}
\Cref{tab:overall} reports the overall performance of \model and \modelplus compared to all baselines across all benchmark datasets. The best, second, and third performing methods are highlighted in bold, underlined, and double-underlined, respectively. 

The overall observation is that \model and \modelplus consistently outperform all baselines across all datasets under both metrics, demonstrating the effectiveness of our proposed techniques for CTA and CPA tasks. \modelplus and \model achieve the average rank of 1.08 and 2.0, respectively, indicating their superior performance compared to the other methods.
Especially on the datasets with wide tables with large average columns per table (\gitdb, \gitsc, \sotabcta, and \sotabcpa), \model and \modelplus achieve significant improvements over the best baselines in \micro and \macro metrics. For example, on \gitdb, \modelplus achieves  {4.05\% and 4.64\%} improvement in \micro and \macro over the best baseline {\qwen (5-shot)}, and \model also achieves significant improvements of {2.31\% and 2.74\%} in \micro and \macro.
On \sotabcpa, \modelplus and \model achieve significant improvements as well. \modelplus improves \macro by 2.46\% and \model by 2.12\% over the best baseline \starmie.
On \wikicta and \wikicpa, where the average number of columns per table is not large, baselines perform well, but \model and \modelplus still outperform all baselines.  
The overall results demonstrate the effectiveness of our proposed retrieval, context-aware encoding, and verification techniques in \cref{sec:retrieval,sec::encoding,sec:verification}, 
to select high-quality context columns for accurate CTA and CPA tasks, especially in handling wide tables.

Moreover, observe that \modelplus consistently outperforms \model across almost all datasets under \micro and \macro, except \wikicta where \model performs slightly better in \micro. The average rank of \modelplus is 1.08. The improvement of \modelplus over \model is particularly significant in \macro, which treats majority and minority classes equally. For example, on \gitdb and \gitsc, \modelplus improves \macro by 1.90\% and 3.84\%, respectively. This indicates the verification model in \cref{sec:verification} is effective in refining the retrieved context columns and improving the performance. 

Besides, among the baselines, \starmie performs the best, with average rank  of 3.5, and it adopts contrastive pretraining to capture column semantics. The baselines designed for annotation tasks, e.g., \watchog and \doduo, typically achieve better performance than the LLM-based methods. This indicates that specific techniques need to be designed for table annotation tasks.  The LLMs yield unstable performance across datasets. \qwen (0-shot) and (5-shot) perform well on \gitdb and \gitsc, but they are significantly worse on the other four datasets, with average ranks of 8.67 and 7.58, respectively; \tllama achieves strong results in \wikicta and \wikicpa but fails in the other four datasets, resulting in rank of 8.92. This inconsistency may depend on if similar table corpus were seen during its pretraining or not. These results suggest that table annotation tasks remain challenging and cannot be totally solved by LLMs alone.

Moreover, we report that \modelplus obtains a subset of columns for 88.91\% of tables in \gitdb, 84.25\% in \gitsc, 93.44\% in \sotabcta, 76.99\% in \sotabcpa, 25.00\% in \wikicta, and 35.81\% in \wikicpa. This demonstrates that our retrieval and verification techniques (Sections 4 and 5) effectively identify relevant context columns for most targets. Although \wikicta and \wikicpa have narrower tables (Table 2), our approach still refines context for a significant portion of tables.

\header\textbf{Performance on Tables with Different Column Numbers.} 
We further analyze the performance of \model and \modelplus on tables with different column numbers, in groups of 1-4, 5-8, 9-12, 13-16, and 17+ columns, as shown in \cref{fig:column_num}.
Observe that the baselines tend to have significant performance drop on tables with many columns, especially on wide tables with more than 16 columns, e.g., \starmie on \gitdb and \gitsc, and \watchog on \wikicta and \wikicpa. This is likely due to the increased complexity and noise introduced by irrelevant or redundant columns, which can hinder the model's ability to focus on the most informative features for annotation tasks. In contrast, \model and \modelplus maintain relatively stable performance when the number of columns increases, especially when the number of columns exceeds 16 on \gitdb, \gitsc, \wikicta and \wikicpa.
Furthermore, in \gitdb and \gitsc, there are 21 and 8 targets in tables with 100+ columns, respectively. In Table \ref{tab:long}, we report the performance  on these extremely wide tables.
 Note that a small sample size may make performance sensitive to incorrect predictions. Nevertheless, \modelplus consistently performs the best, followed by \model.
The observations validate the proposed techniques in \cref{sec:retrieval,sec:verification}, to explicitly select the most informative context columns for CTA and CPA tasks, which helps mitigate the noise introduced by irrelevant columns. This leads to more robust performance of \modelplus and \model across different table sizes. 
\begin{table}[tp]
  \centering
  \caption{Accuracy (\%) on tables with more than 100 columns.}
  \label{tab:long}
  \vspace{-4mm}
  \resizebox{0.88\linewidth}{!}{
  \renewcommand{\arraystretch}{0.92}
  \begin{tabular}{lccccc}
    \toprule
    {Dataset} & {\starmie
    } & {\reca} & {\watchog
    }  & {\model} & {\modelplus}  \\
    \midrule
    \gitdb  & 28.6 & 47.6 & 42.7 & 52.4 & \textbf{66.7} \\
    \gitsc   & 25.0 & \textbf{62.5} & 37.5 & \textbf{62.5} & \textbf{62.5} \\
    \bottomrule
  \end{tabular}
  }
  \vspace{-0mm}
\end{table}

\subsection{Efficiency Analysis}\label{sec::efficiency} 
\begin{figure}[tp]
    \centering
      \includegraphics[width=0.48\textwidth]{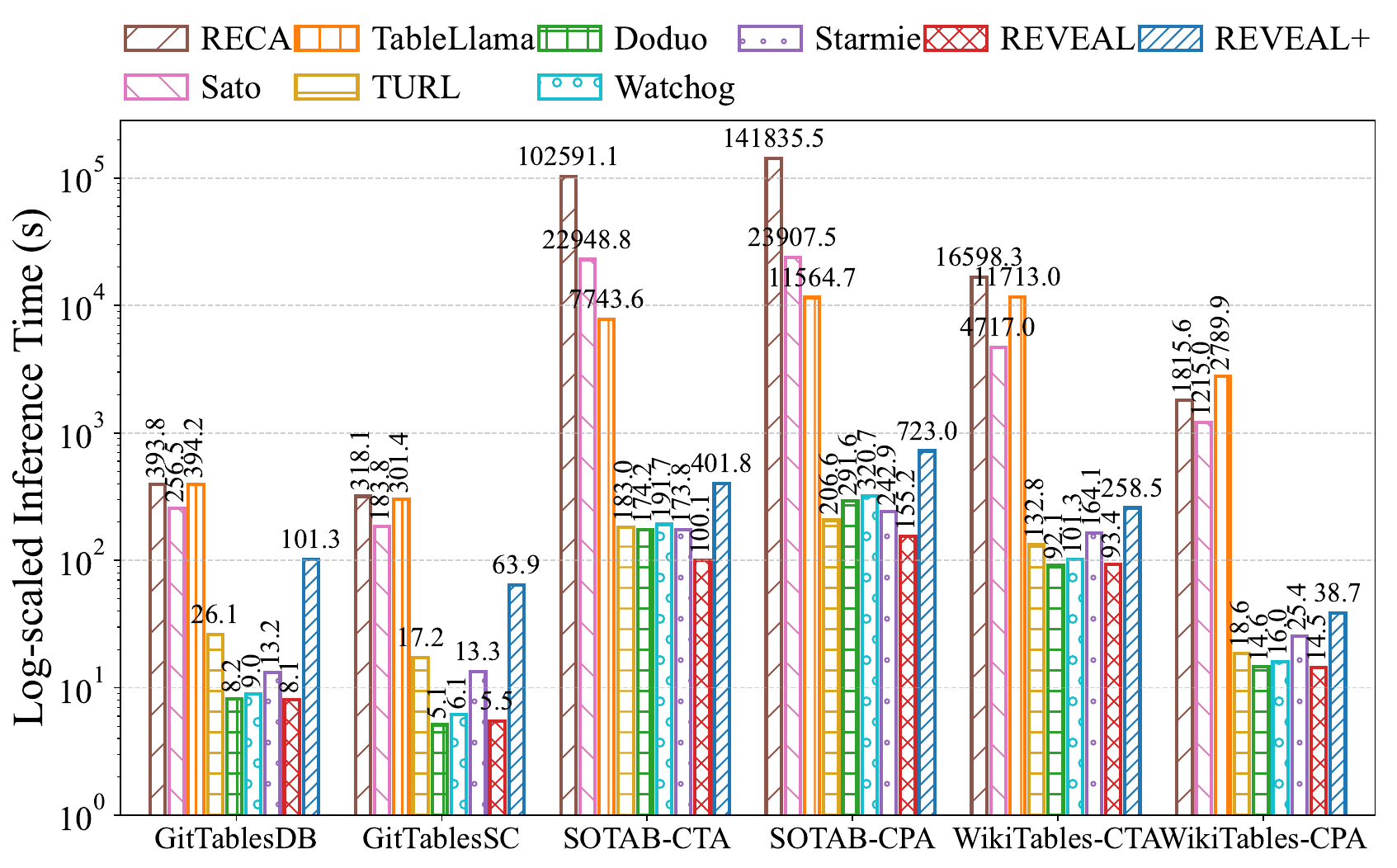}
    \caption{Inference time comparison.}
\label{fig:inference_time}
    \vspace{-1.5mm}

\end{figure}

\cref{fig:inference_time} reports the inference time of \model and \modelplus compared to other baselines.  
First, observe that \model ranks top among the methods across all datasets, with the fastest inference time on \sotabcta and \sotabcpa and similar efficiency as \doduo on the other datasets. 
The efficiency of \model is attributed to its retrieval-based approach, which avoids the need for full-table input. 
More importantly, as shown in \cref{tab:overall}, \model achieves better quality than existing methods for the annotation tasks. 
This indicates that \model can achieve high-quality annotation with a more efficient inference process. 
Second, \modelplus incurs additional inference time due to running the verification model to further refine the column context, in order to further boost effectiveness over \model. But still, the efficiency of \modelplus is moderate among all methods, substantially faster than \reca, \sato and \tllama. As shown in \cref{tab:overall}, \modelplus achieves the top-1 performance on almost all settings, significantly improving the performance of \model. 
As mentioned, in the CTA and CPA tasks, result quality is relatively more important since the inference can be done offline and time is not a critical factor. 
Therefore, \modelplus serves as a good trade-off for effectiveness over efficiency.

\begin{figure}[tp]
    \centering
      \includegraphics[width=0.45\textwidth]{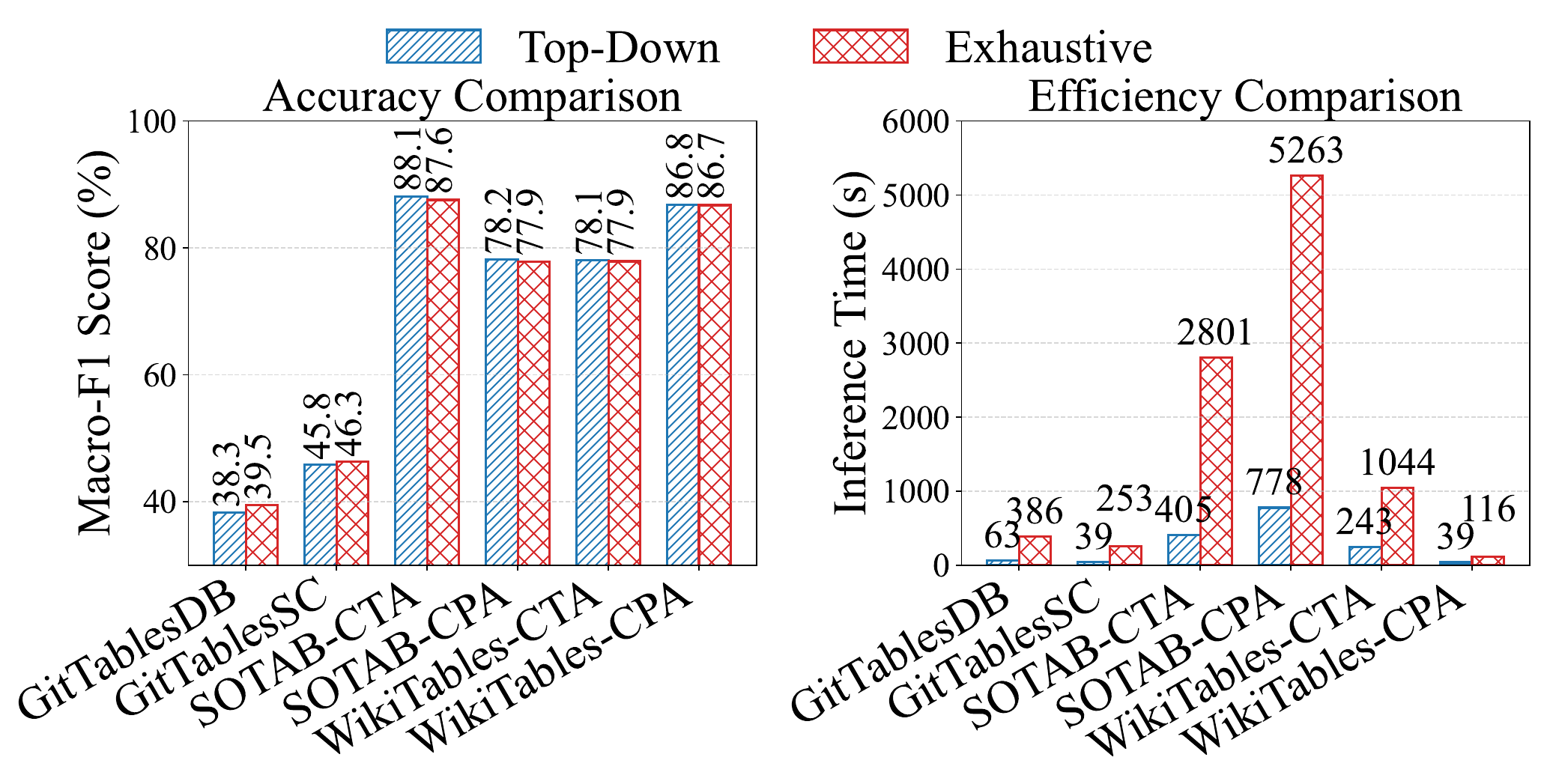}
    \caption{Comparison between the Top-Down inference strategy and Exhaustive search in terms of accuracy (\macro) and inference efficiency.}
    \label{fig:topdown}
        \vspace{-3mm}
\end{figure}

\header
\textbf{Top-Down Verification Inference vs. Exhaustive Verification.}
We evaluate the greedy top-down verification inference designed in \cref{sec::inference} against exhaustive verification, which evaluates all possible context subsets $\vcan$ to get $\vpool$, in terms of both accuracy and efficiency, in \cref{fig:topdown}. 
The left figure shows that the top-down strategy achieves nearly the same Macro-F1 as exhaustive search, indicating that it effectively finds the high-quality verified column context. 
The right figure shows that the top-down verification inference is significantly more efficient than exhaustive search. For example, on \sotabcpa, the top-down strategy reduces inference time from 5263s to 778s, achieving over 6x speedup while maintaining accuracy.
\cref{fig:topdown} demonstrates the effectiveness and efficiency of the proposed top-down verification inference technique.

\subsection{Experimental Analysis}\label{sec::sensitivity}

\begin{figure}[tp]
    \centering
      \includegraphics[width=0.36\textwidth]{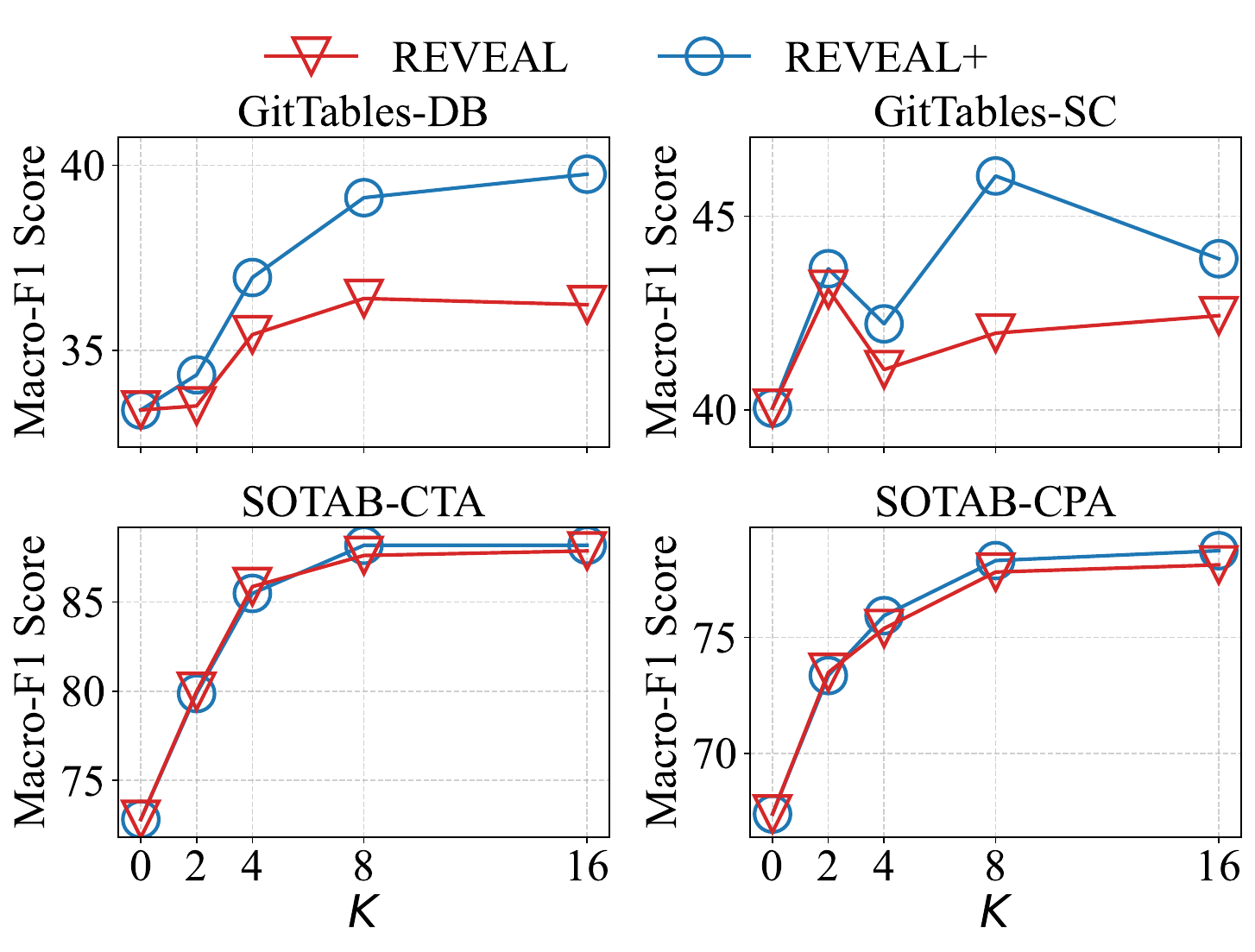}
      \vspace{-1mm}
    \caption{Vary $K$.}
      \vspace{0.5mm}

    \label{fig:exp_size}
\end{figure}

\header
\textbf{Varying $K$ of Column Context Size.}
In the retrieval method, we retrieve a column context of $K$ columns for a target in a table. 
We evaluate the impact of varying $K$ in {\{0, 2, 4, 8, 16\}}  on the performance of \model and \modelplus, as reported in \cref{fig:exp_size}.
Observe that the performance of both \model and \modelplus improves as $K$ increases to 8 on all datasets, except \gitsc with a little fluctuation. 
Then the performance becomes stable with further increase of $K$ to 16. 
Observe that at the same $K$ value, \modelplus outperforms \model, demonstrating the effectiveness of the verification techniques in Section~\ref{sec:verification}. \cref{fig:exp_size} shows that using few columns with a small $K$ may omit critical context, while increasing $K$ beyond a moderate value yields diminishing returns. Therefore, we set $K=8$ by default.

 \begin{table}[tb]
  \centering
  \caption{Ablation study   in \macro (\%).}
  \vspace{-4mm}
    \setlength{\tabcolsep}{3pt}
   \resizebox{0.98\linewidth}{!}{
   \renewcommand{\arraystretch}{1}
    \begin{tabular}{ccccc}
    \toprule
          & \multicolumn{1}{l}{\gitdb} & \multicolumn{1}{l}{\gitsc} & \multicolumn{1}{l}{SOTAB-CTA} & \multicolumn{1}{l}{SOTAB-CPA} \\
    \midrule
    w/o encoding & 30.69  & 39.22  & 87.05  & 77.48  \\
    TURL encoding & 27.22  & 37.98  & 86.89  & {76.98} \\
    context-aware encoding & \textbf{36.40} & \textbf{41.98} & \textbf{87.60} & \textbf{77.81} \\
    \bottomrule
    \end{tabular}%
    }
  \label{tab:ablation}
  \vspace{-2mm}
\end{table}

\header 
\textbf{Ablation on Context-aware Encoding.}  
We conduct an ablation study on the context-aware encoding in Section~\ref{sec::encoding} by comparing \model with and without the encoding, as well as with TURL encoding~\cite{turl}. In \cref{tab:ablation}, our proposed context-aware encoding achieves the best performance across all datasets. For example, on \gitdb, \model with context-aware encoding improves \macro from 30.69\% (without encoding) to 36.40\% and also outperforms TURL encoding by a large margin.  These results demonstrate the effectiveness of distinguishing context columns from target columns during encoding, as designed in Section~\ref{sec::encoding}.

\begin{table}[t!]
  \centering
\caption{Retrieval method vs. other strategies.}
\vspace{-3.5mm}
  \setlength{\tabcolsep}{3pt}
   \resizebox{0.98\linewidth}{!}{
    \renewcommand{\arraystretch}{0.92} 
    \begin{tabular}{ccccc|c}
    \toprule
     \micro & \multicolumn{1}{c}{\gitdb} & \multicolumn{1}{c}{\gitsc} & \multicolumn{1}{c}{\sotabcta} & \multicolumn{1}{c}{\sotabcpa} & \multicolumn{1}{|c}{Avg. Rank}  \\
    \midrule
    Random & 58.57  & 67.52  & 88.02  & 79.56  & {4.75}  \\
    First & 60.13  & 68.25  & 87.55  & 79.68  & {4.25}  \\
    Nearby & 60.18  & 68.85  & \underline{88.28}  & 80.78  & {2.75}  \\
    Similar & 58.47  & 67.86  & 87.87  & \textbf{81.19 } & {4.0}  \\
    Position & \underline{60.40} & \underline{69.24} & 87.51 & 78.07 & 4.0 \\
Ours & \textbf{61.53} & \textbf{70.9} & \textbf{88.74} & \underline{80.81} & {\textbf{1.25}} \\
    \bottomrule
    \end{tabular}%
    }
  \label{tab:retrieval}
      \vspace{0.5mm}

\end{table}%

\header 
\textbf{Study on the Retrieval Technique in {\cref{sec:retrieval}}.}
In \cref{sec:retrieval}, we propose a column retrieval method   that selects a compact and informative subset of $K$ columns as the column context $\candpool$. Our design emphasizes both semantic relevance and diversity, and employs MMR for context selection.
We compare it with several alternative strategies. In particular, we compare the following strategies: \textit{Random} selects $K$ context columns at random; \textit{First} takes the first $K$ columns from the table; \textit{Nearby} selects the columns adjacent to the target; \textit{Similar} retrieves the top-$K$ columns with the highest embedding similarity to the target column; \textit{Position} selects the two leftmost columns and the left and right columns of the target. Note that all these methods are implemented within the same \modelplus framework and only differ in how to get $\candpool$ in \cref{sec:retrieval}. 
The results are reported in \cref{tab:retrieval}, and the improvement of ours in the last row over the others is solely achieved by \cref{sec:retrieval}. 
Observe that our method with the retrieval technique in \cref{sec:retrieval} performs better than the other strategies, except \textit{Similar} on \sotabcpa. 
This confirms the effectiveness of combining semantic relevance and diversity in column retrieval, as emphasized in our design.

\begin{figure}[tp]
    \centering
      \includegraphics[width=0.32\textwidth]{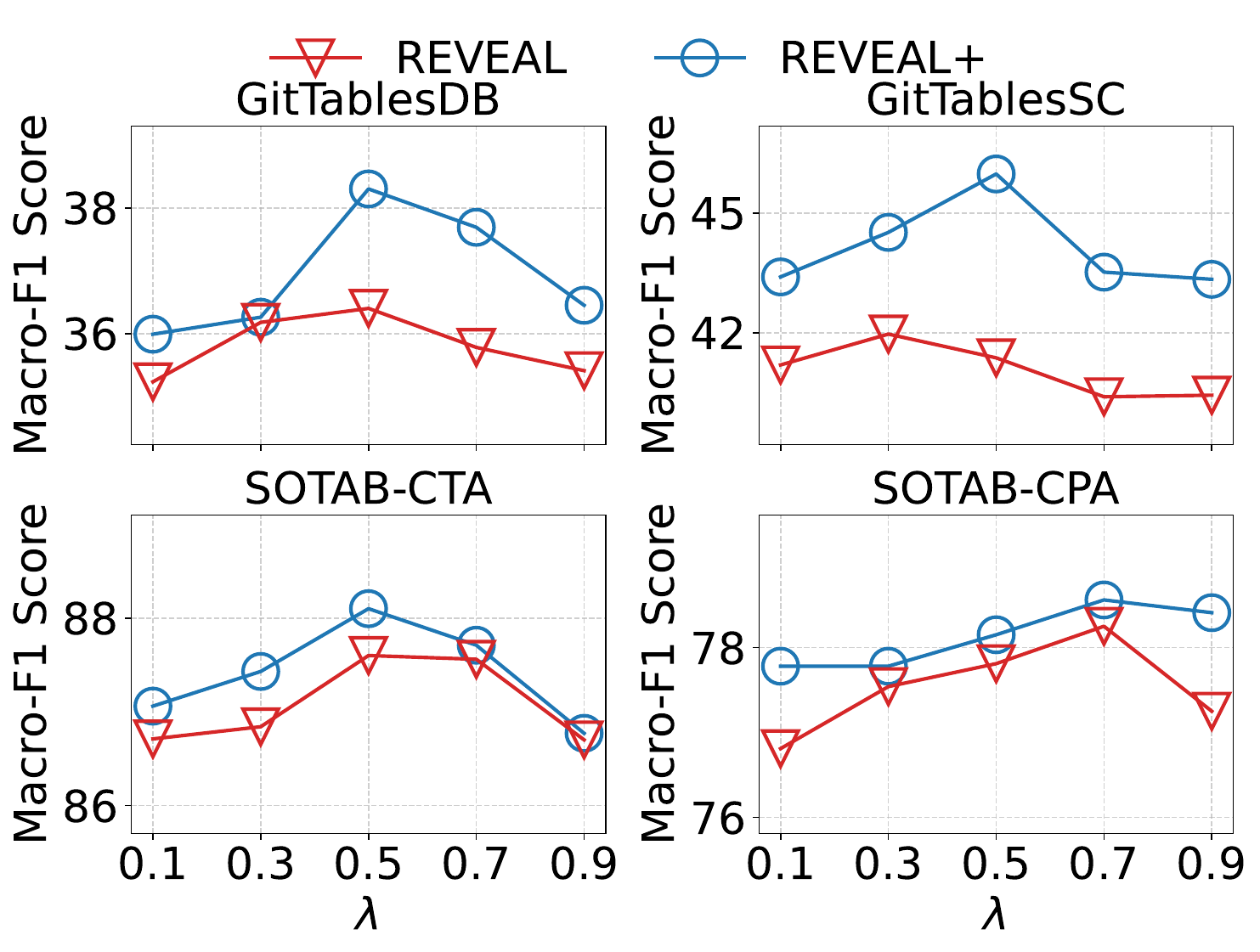}
      \vspace{-1mm}
    \caption{Vary $\lambda$.}    \label{fig:lbd}

      \vspace{-0.5mm} 

\end{figure}

\header 
\textbf{Study on \cref{eq:candpoolGeneration}.}
We study Equation (2) by introducing a weight $\lambda$ to control the trade-off between relevance and diversity: $\mr(c,\target, \candpool) = {\lambda} \operatorname{cos}(\embcol, \mathbf{e}_{\target}) - {(1-\lambda)} \max_{c'' \in \candpool } \operatorname{cos}(\embcol, \mathbf{e}_{c''})$.  
A smaller $\lambda$ emphasizes diversity, while a larger $\lambda$ emphasizes relevance. By default, we consider same weight for both terms, i.e.,  $\lambda=0.5$. 
\cref{fig:lbd} reports the performance of \model and \modelplus for $\lambda$ in $\{0.1, 0.3, 0.5, 0.7, 0.9\}$. 
As $\lambda$ increases, the performance of both methods generally improves and then plateaus or decreases. 
\modelplus consistently outperforms \model across all datasets. Notably, \modelplus achieves its best performance at $\lambda=0.5$ on most datasets, except for \sotabcpa, where \modelplus with $\lambda=0.7$ yields even higher quality.
These results indicate that it is not possible to simply tune~\cref{eq:candpoolGeneration} to let \model to achieve the same performance as \modelplus, validating the necessity of the verification techniques in Section~\ref{sec:verification} for \modelplus to further  enhance performance.

\begin{figure}[tp]
    \centering
      \includegraphics[width=0.41\textwidth]{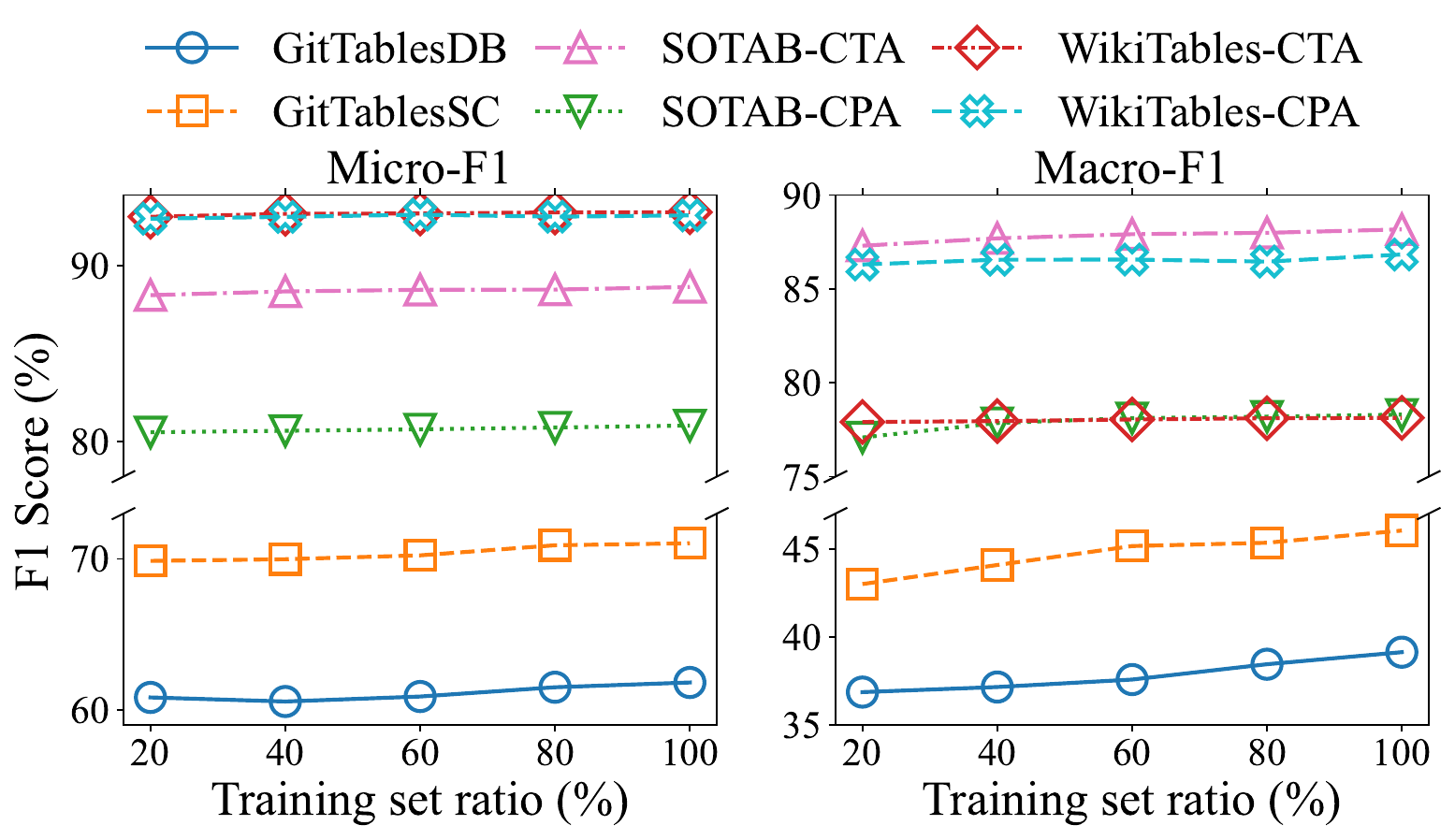}
    \caption{Learning efficiency of the verification model in \modelplus by varying the ratio of training data.}
    \label{fig:frac}
        \vspace{-1.5mm}
\end{figure}

\header
\textbf{Learning Efficiency of the Verification Model.} 
In \cref{sec:verification}, we construct a labeled dataset to train the verification model. To evaluate its learning efficiency, we vary the size of its training set from 20\% to 100\% and report the results of \modelplus in \cref{fig:frac}.
When the training set ratio increases, the performance of \modelplus generally improves in a mild way, indicating that the verification model benefits from more training data, but is not sensitive to that and can achieve good performance with limited training data.
Note that on \gitdb and \gitsc, the performance of improves more significantly than on the other datasets. The reason is that \gitdb and \gitsc contain more complex tables with wider structures. Therefore, we need to have a sufficiently trained verification model to verify if a subset of columns is helpful for the target annotation tasks. 
\cref{fig:frac} illustrates that the verification model with the top-down inference technique in \cref{sec:verification} is effective and robust to learn from training data.

\begin{table}[t!]
  \centering
\caption{\modelplus with different verification methods.
}
\vspace{-3.5mm}
  \resizebox{\linewidth}{!}{
               \renewcommand{\arraystretch}{0.92}
    \begin{tabular}{ccccc}
    \toprule
        \macro  & \multicolumn{1}{c}{\gitdb} & \multicolumn{1}{c}{\gitsc} & \multicolumn{1}{c}{\sotabcta} & \multicolumn{1}{c}{\sotabcpa} \\
    \midrule
    Random & 34.96 & 38.31 & 76.31 & 67.90 \\
    Max Confidence & 34.83 & 34.32 & 76.35 & 69.77 \\
    Majority Voting & 36.71 & 40.02 & 83.69 & 73.08  \\
    Weighted Voting & 36.73  & 39.27  & 84.07  & 73.77  \\
    \modelplus & \textbf{38.30} & \textbf{45.82} & \textbf{88.1} & \textbf{78.15} \\
    \bottomrule
    \end{tabular}%
    }
        \vspace{1mm}
  \label{tab:verify}
\end{table}%

\header
\textbf{Study on the Verification Model.}  
In \modelplus, we employ a learned verification model with a top-down inference strategy to identify $\vpool$ from $\candpool$. We compare it against several alternative methods: \textit{Random}, which selects $\vpool$ from $\candpool$ randomly, with results averaged over 10 runs; \textit{Max Confidence}, which selects the subset with the highest softmax probability from the prediction module $\tmodel$; \textit{Majority Voting}, which determines the final label based on the most frequent label predicted across all subsets of $\candpool$; and \textit{Weighted Voting}, which averages the softmax probabilities across all subsets and selects the label with the highest mean probability.
\cref{tab:verify} reports the \macro results. Observe that \modelplus with the proposed verification model consistently outperforms all other methods across all datasets, with a significant margin. 
This demonstrates the effectiveness of our verification model in selecting informative column contexts for the target column, leading to improved performance in the annotation tasks.

\begin{table}[t]
  \centering
  \caption{Performance of baselines with $\candpool$.}
        \vspace{-4mm}
     \resizebox{1\linewidth}{!}{
       \renewcommand{\arraystretch}{0.85}
    \begin{tabular}{ccccccccc}
    \toprule
          & \multicolumn{2}{c}{\gitdb} & \multicolumn{2}{c}{\gitsc} & \multicolumn{2}{c}{SOTAB-CTA} & \multicolumn{2}{c}{SOTAB-CPA} \\
    \midrule
    \watchog & 53.96  & 28.89  & 65.24  & 36.06  & 86.23  & 84.19  & 76.52  & 72.60  \\
    \watchog + $\candpool$ & 54.27  & 29.32  & 65.35  & 37.06  & 87.90  & 86.47  & 79.66  & 75.91  \\
    \midrule
    \starmie & 54.43  & 30.71  & 64.87  & 37.04  & 87.95  & 86.84  & 78.92  & 75.69  \\
    \starmie + $\candpool$ & 56.98  & 32.11  & 66.88  & 37.47  & 88.29  & 87.49  & 80.40  & 76.78  \\
    \midrule
    \modelplus &    \textbf{61.53} & \textbf{38.3} & \textbf{70.9} & \textbf{45.82} & \textbf{88.74} & \textbf{88.1} & \textbf{80.81} & \textbf{78.15} \\
    \bottomrule
    \end{tabular}%
    }
  \label{tab:baselineC}%
  \vspace{-2mm}
\end{table}%

\header
\textbf{Applying Baselines over $\candpool$.}  We apply strong baselines, \watchog and \starmie, on the column context $\candpool$ retrieved in \cref{sec:retrieval}.   As shown in \cref{tab:baselineC}, both baselines exhibit improved performance when using $\candpool$, compared to using original tables. 
This demonstrates the effectiveness and generalizability of our idea of retrieving a compact and informative column context for the target. Nonetheless, \modelplus still significantly outperforms both baselines,  when they use the same $\candpool$, validating the effectiveness and necessity of our techniques in~\cref{sec::encoding} and \cref{sec:verification}.

\begin{table}[t] 
  \centering
  \caption{Top-down vs.\ bottom-up verification inference.}
  \vspace{-4mm}
     \resizebox{0.95\linewidth}{!}{
           \setlength{\tabcolsep}{1pt}
           \renewcommand{\arraystretch}{0.92}

    \begin{tabular}{lcccccccc}
    \toprule
          & \multicolumn{2}{c}{\gitdb} & \multicolumn{2}{c}{\gitsc} & \multicolumn{2}{c}{\sotabcta} & \multicolumn{2}{c}{\sotabcpa} \\
    \midrule
          & \micro & \macro & \micro & \macro & \micro & \macro & \micro & \macro \\
    \midrule
    bottom-up & 55.86  & 31.09  & 63.49  & 28.48  & 65.29  & 62.30 & 68.43 & 64.85  \\
    top-down & \textbf{61.53} & \textbf{38.30} & \textbf{70.90} & \textbf{45.82} & \textbf{88.74} & \textbf{88.10} & \textbf{80.81} & \textbf{78.15} \\
    \bottomrule
    \end{tabular}
    }
  \label{tab:bottom}%
    \vspace{-2mm}
\end{table}%

\header
\textbf{Top-down vs. Bottom-up Verification.} 
We compare the top-down verificaiton inference in~\cref{sec::inference} with a bottom-up approach. 
The bottom-up approach starts with the top-ranked column in \cref{alg:retrieval} and iteratively adds columns until either all columns in $\candpool$ are included or the quality score from the verification model no longer improves, i.e., early stopping. Table~\ref{tab:bottom} shows that bottom-up search is outperformed by our top-down approach.  
This supports our intuition in~\cref{sec::inference} that top-down verification, which starts from a larger high-quality context and prunes only a few noisy columns, can get better $\vpool$. In contrast, bottom-up verification may be more sensitive to the initial column choices.

\begin{table}[t]
  \centering
  \caption{ \modelplus using  $\candpool$ vs. full column set.}
  \vspace{-4mm}
    \resizebox{0.96\linewidth}{!}{
    \renewcommand{\arraystretch}{0.94}
      \setlength{\tabcolsep}{1pt}
    \begin{tabular}{ccccccccc}
    \toprule
          & \multicolumn{2}{c}{\gitdb} & \multicolumn{2}{c}{\gitsc} & \multicolumn{2}{c}{SOTAB-CTA} & \multicolumn{2}{c}{SOTAB-CPA} \\
    \midrule
          & \multicolumn{1}{l}{\micro} & \multicolumn{1}{l}{\macro} & \multicolumn{1}{l}{\micro} & \multicolumn{1}{l}{\macro} & \multicolumn{1}{l}{\micro} & \multicolumn{1}{l}{\macro} & \multicolumn{1}{l}{\micro} & \multicolumn{1}{l}{\macro} \\
    \midrule
    full set & 60.27  & 36.75  & 69.58  & 42.45  & 88.37  & 87.61  & 79.60  & 77.11  \\
    $\candpool$   & {\textbf{61.80}} & {\textbf{39.12}} & {\textbf{71.03}} & {\textbf{46.04}} & {\textbf{88.78}} & {\textbf{88.17}} & {\textbf{80.90}} & {\textbf{78.30}} \\
    \bottomrule
    \end{tabular}
    } 
  \label{tab:fullset}%
    \vspace{1mm}
\end{table}%
\header
\textbf{Performance of \modelplus on Full Column Set.} 
We compare  \modelplus when verifying over the retrieved context $\candpool$ versus verifying over the full set of columns in a table. As shown in \cref{tab:fullset}, \modelplus achieves consistently better results using $\candpool$ than using the full column set. 
This validates that the retrieval stage in \model effectively eliminates irrelevant columns, enabling more focused and informative context for effective verification in \modelplus.

\begin{table}[t]
  \centering
  \caption{Performance on different types of columns.}
  \vspace{-4mm}
  \resizebox{0.86\linewidth}{!}{
  \renewcommand{\arraystretch}{0.92}
  \begin{tabular}{ccccccc}
    \toprule
          & \multicolumn{3}{c}{\gitdb} & \multicolumn{3}{c}{\gitsc} \\
    \midrule
          & text  & numeric & date-time & text  & numeric & date-time \\
    \midrule
    \starmie & 49.85  & 70.51  & 53.01  & 56.62  & 80.54  & 58.00  \\
    \modelplus & \textbf{56.88} & \textbf{70.78} & \textbf{59.04} & \textbf{61.76} & \textbf{84.90} & \textbf{62.00} \\
    \bottomrule
    \end{tabular}
    }
  \label{tab:type}%
    \vspace{-2mm}
\end{table}%
\header
\textbf{Performance on Different Column Types.}
We analyze performance by column type: text (e.g., name, description, title), numeric (e.g., value, age, price), and datetime (e.g., date, year, time). 
Table~\ref{tab:type} reports the results of \modelplus and the baseline \starmie for each column type. 
Both methods achieve higher accuracy on numeric columns, likely because numeric types are more distinct and easier to classify than text or datetime columns. Importantly, \modelplus consistently outperforms \starmie across all types.

\header
\textbf{Performance with 512 Tokens.} We set the maximum input length of BERT to 256 tokens, following~\cite{watchog,starmie}.  In~\cite{reca}, it shows that varying between 256 and 512 tokens does not affect performance much. In Table~\ref{tab:512},  we report the results with 512 tokens, comparing with strong baselines.  
Observe that using 512 tokens yields minor improvements over 256 tokens, and our \modelplus and \model consistently outperform the baselines.
These results further demonstrate the effectiveness and robustness of our methods.

\begin{table}[t]
  \centering
  \caption{Results with 512 tokens.}
  \vspace{-4mm}
  \renewcommand{\arraystretch}{0.92}
  \resizebox{1\linewidth}{!}{
    \setlength{\tabcolsep}{1pt}
    \begin{tabular}{ccccccccc}
    \toprule
          & \multicolumn{2}{c}{GittablesDB} & \multicolumn{2}{c}{GittablesSC} & \multicolumn{2}{c}{SOTAB-CTA} & \multicolumn{2}{c}{SOTAB-CPA} \\
          & \micro & \macro & \micro & \macro & \micro & \macro & \micro & \macro \\
    \midrule
    \watchog & 54.27  & 29.37  & 64.44  & 34.75  & 86.56  & 84.57  & 78.06  & 71.90  \\
    \starmie & 54.16  & 31.44  & 64.85  & 38.31  & 87.92  & 86.91  & 78.77  & 75.77  \\
    \model & \underline{59.88} & \underline{36.69} & \underline{70.33} & \underline{42.41} & \underline{89.40} & \underline{88.66} & \underline{80.68} & \underline{78.05} \\
    \modelplus & \textbf{62.07} & \textbf{39.56} & \textbf{71.36} & \textbf{46.35} & \textbf{89.79} & \textbf{89.18} & \textbf{80.99} & \textbf{78.16} \\
    \bottomrule
    \end{tabular}%
    }
  \label{tab:512}%
  \vspace{-2mm}

\end{table}%

\begin{figure}[t]
    \centering
      \includegraphics[width=0.49\textwidth]{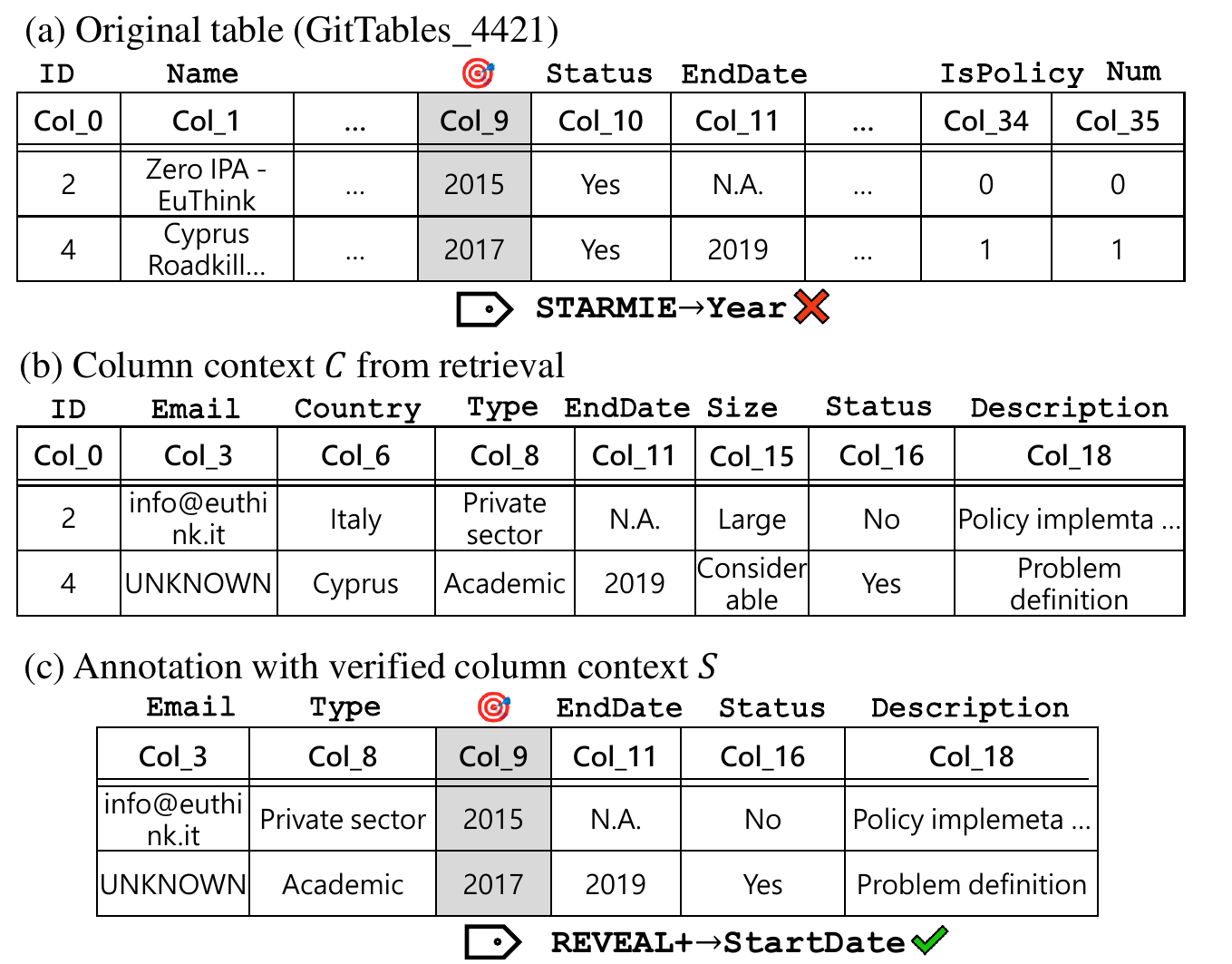}
    \caption{Illustrative Example}
    \label{fig:case}
    \vspace{1mm}
\end{figure}

\subsection{Illustrative Example}\label{sec::case}
\cref{fig:case} presents an example on a table GitTables\_4421 with 136 columns from \gitdb, describing citizen science projects.  
\cref{fig:case}(a) shows a sample view of the table, where the semantic types of columns are provided for reference but are not used in the CTA task. 
The target column Col\_9 in gray contains year-like values, such as \texttt{2015} and \texttt{2017}. Its ground truth type is \texttt{StartDate}, indicating the start date of a project, paired with the \texttt{EndDate} column Col\_11. 
The strong baseline, \starmie, predicts the target column as \texttt{Year} when using the entire table as input. While reasonable, this prediction is incorrect. 
The table contains many irrelevant columns, such as Col\_34 and Col\_35, which do not contribute to predicting the target column type. These irrelevant columns overwhelm the baseline model, leading to suboptimal predictions.
\cref{fig:case}(b) shows the column context $\candpool$ with 8 columns retrieved using our retrieval method in \cref{sec:retrieval}. These columns, such as \texttt{EndDate}, \texttt{Status}, and \texttt{Description}, are closely related to the project information and help clarify the target column's meaning. This demonstrates the retrieval method's ability to filter out irrelevant columns effectively.
\cref{fig:case}(c) shows the verified column context $\vpool$ selected from $\candpool$ by the verification model in \modelplus (\cref{sec:verification}). Compared to $\candpool$, the verification model further refines the context by removing less relevant columns, such as \texttt{Size}, ensuring only the most informative columns remain. 
With the verified column context, \modelplus correctly predicts the target column type as \texttt{StartDate}, matching the ground truth and together with the \texttt{EndDate} column, indicating a complete project timeline.

\section{Related Work}
\label{sec:related}

Early methods for column annotation rely on hand-crafted and statistical data features.  SemanticTyper~\cite{SemanticTyper} utilized TF-IDF for textual data and Kolmogorov-Smirnov tests~\cite{lehmann2005testing} for numeric data to distinguish data types. Pham et al.~\cite{pham2016} extended this approach by incorporating additional features, such as the Mann-Whitney test for numerical data and Jaccard similarity for textual data, to train logistic regression and random forest models for  annotation.

Building on advancements in machine learning, recent studies have incorporated semantic features, framing CTA and CPA as multi-class classification problems. Sherlock~\cite{sherlock} extracts multi-granularity tabular features, such as character-level, word-level, segment-level, and global-level, and trains deep learning classifiers for semantic type prediction. Sato~\cite{sato} extends Sherlock by modeling table-level topics and correlations between neighbor columns. 

Subsequently, language models like BERT~\cite{bert} have been employed to learn representations of tabular data for table understanding~\cite{tabert,tabbie}. Column annotation methods based on language models~\cite{turl,doduo,watchog} have gained significant attraction. For instance, TURL~\cite{turl} introduces a pre-training and fine-tuning framework that uses a visibility matrix to mask irrelevant table components and generate column embeddings for downstream tasks. \doduo~\cite{doduo} developes a multi-task learning framework based on BERT, which takes the entire table as input and predicts column types and relations using a single model. It achieves high annotation quality by modeling token-level interactions across columns through self-attention. 
RECA~\cite{reca} aligns schema-similar and topic-related tables with a novel named entity schema to address the complexities of wide tables and inter-table contexts. Watchog~\cite{watchog} employs contrastive learning techniques to tackle challenges associated with data sparsity and class imbalance in column annotation tasks, reducing reliance on high-quality annotated instances. 
Starmie~\cite{starmie}, originally designed for dataset discovery in data lakes, proposes a self-supervised contrastive learning framework to train a high-quality column encoder. We adapt it to column annotation and compare its performance with our methods.
These studies underscore the importance of effective representational learning for accurate annotations. However, existing models typically process all columns in a table as input for a target, relying on transformers and attention mechanisms to infer useful semantics and column interactions. This approach often fails to filter out irrelevant columns, introducing noise that can degrade performance. In contrast, our method explicitly identifies and validates contextually relevant columns for the target column, enabling a more precise understanding of column semantics. 
\citet{DBLP:journals/pvldb/ShragaM23} propose a different problem, semantic data versioning, focusing on explaining changes between dataset versions by identifying transformations from an origin relation to a goal relation. Their Explain-Da-V method uses functional dependency (FD) discovery to select column subsets for efficient search. In contrast, we work on column annotation for tables with missing metadata—a distinct task that could benefit data versioning. Our retrieval and verification techniques in Sections \ref{sec:reveal} and \ref{sec:verification} are technically different from the FD-based approach in~\cite{DBLP:journals/pvldb/ShragaM23}.

Recent studies have explored the use of LLMs~\cite{achiam2023gpt,llama3,DBLP:journals/corr/abs-2407-10671} for column annotation tasks~\cite{tablellama,archetype,tablegpt}. For instance, \tllama~\cite{tablellama} fine-tunes LLMs on various table-related tasks, including column annotation. In our experiments, we compared our method with \tllama and a general-purpose LLM, Qwen-Plus~\cite{DBLP:journals/corr/abs-2407-10671}, and observed that they perform suboptimally on column annotation tasks. This suggests that the effectiveness of LLMs in such tasks depends heavily on whether they have been trained or fine-tuned on relevant table-specific corpora. Consequently, column annotation remains a challenging problem that cannot be effectively addressed by LLMs alone and requires dedicated designs and methods.

In addition, several other tasks are related to tabular data, including table discovery~\cite{DBLP:conf/sigmod/Fan00M23,DBLP:conf/hilda/HulsebosLSP24,DBLP:journals/pvldb/DengCCYCYSWLCJZJZWYWT24}, such as table join search~\cite{DBLP:conf/sigmod/ZhuDNM19,DBLP:conf/icde/DongT0O21,DBLP:journals/pvldb/Dong0NEO23} and table union search~\cite{DBLP:journals/pvldb/NargesianZPM18,santos,starmie}, as well as schema matching~\cite{DBLP:journals/pacmmod/TuFTWL0JG23,DBLP:conf/icde/KoutrasSIPBFLBK21} and tabular data synthesis for dataset augmentation~\cite{Tenet,Privlava}. These tasks rely heavily on effective table understanding and column representation learning. In future work, we plan to extend our techniques to support these broader table-related applications.

\section{Conclusion}\label{sec:conclusion}
We propose a novel retrieve-and-verify framework comprising the \model and \modelplus, which selectively incorporate relevant column context to enhance annotation accuracy. \model employs an unsupervised retrieval strategy to construct compact and informative column subsets, combined with context-aware encoding techniques that differentiate between target and context columns to learn effective embeddings. Building on this, \modelplus refines the retrieved column context using a lightweight verification model, formulating context verification as a supervised classification problem and introducing a top-down inference method to efficiently identify high-quality contexts.
Extensive experiments on six benchmark datasets validate the effectiveness of our framework, with both \model and \modelplus significantly surpassing existing state-of-the-art methods. These findings underscore the importance of selective context in table understanding and present a scalable, generalizable solution for real-world applications.
As future work, a direction is to leverage the context quality scores learned by the verification model in \modelplus to guide the retrieval in \model in a supervised manner, potentially improving performance. We also plan to extend our framework to other tabular data tasks, e.g., table generation, and explore integration with advanced large models.


\balance

\bibliographystyle{ACM-Reference-Format}
\bibliography{sample}



\end{document}